\documentclass[lettersize,journal]{IEEEtran}
\IEEEoverridecommandlockouts
\usepackage{cite}
\usepackage{amsmath,amssymb,amsfonts}
\usepackage{graphicx}
\usepackage{textcomp}
\usepackage{xcolor}
\usepackage{tabularx}
\usepackage{subfigure}
\usepackage{colortbl}
\usepackage{algorithm}
\usepackage{algpseudocode}
\usepackage{color}
\usepackage{multirow}
\usepackage{subcaption}
\usepackage{booktabs}
\usepackage{url}

\def\BibTeX{{\rm B\kern-.05em{\sc i\kern-.025em b}\kern-.08em
    T\kern-.1667em\lower.7ex\hbox{E}\kern-.125emX}}
\begin{document}

\title{Denoising Refinement Diffusion Models for Simultaneous Generation of Multi-scale Mobile Traffic\\
}




\author{
    Xiaoqian Qi, Haoye Chai,~\IEEEmembership{Member,~IEEE}, Sichang Liu, Lei Yue, Raoyuan Pan, \\Yue Wang, and Yong Li,~\IEEEmembership{Member,~IEEE}
    
    \thanks{X. Qi, S. Liu, Y. Wang, and Y. Li are with the Department of Electronic Engineering, BNRist, Tsinghua University, Beijing 100084, China. H. Chai is with the State Key Laboratory of Networking and Switching Technology, Beijing University of Posts and Telecommunications, Beijing 100876, China. L. Yue and R. Pan are with China Mobile Communications Group Guangxi Co., Ltd.‌, Nanning 530022, China. E-mail: haoyechai@bupt.edu.cn}
    
}

\markboth{IEEE TRANSACTIONS ON MOBILE COMPUTING}%
{Shell \MakeLowercase{\textit{et al.}}: A Sample Article Using IEEEtran.cls for IEEE Journals}

\maketitle

\begin{abstract} 
The planning, management, and resource scheduling of cellular mobile networks require joint estimation of mobile traffic across different layers and nodes. Mobile traffic generation can proactively anticipate user demands and capture the dynamics of network load. However, existing methods mainly focus on generating traffic at a single spatiotemporal resolution, making it difficult to jointly model multi-scale traffic patterns. In this paper, we propose \textbf{ZoomDiff}, a diffusion-based model for multi-scale mobile traffic generation. ZoomDiff maps urban environmental context into mobile traffic with multiple spatial and temporal resolutions through a set of customized Denoising Refinement Diffusion Models (DRDM). DRDM employs a multi-stage noise-adding and denoising mechanism, enabling different stages to generate traffic at distinct spatiotemporal resolutions. This design aligns the progressive denoising process with hierarchical network layers, including base stations, cells, and grids of varying granularities. Experiments on real-world mobile traffic datasets show that ZoomDiff achieves at least an 18.4\% improvement over state-of-the-art baselines in multi-scale traffic generation tasks. Moreover, ZoomDiff demonstrates strong efficiency and cross-city generalization, highlighting its potential as a powerful generative framework for modeling multi-scale mobile network dynamics. 
\end{abstract}


\section{Introduction} 
Mobile traffic serves as a vital indicator of both user behavior and network performance~\cite{10646623}. Accurate modeling and understanding of traffic dynamics not only reflect underlying communication activities but also enable informed decisions in network planning and optimization. However, the privacy of communication data and the limited availability of large-scale, high-quality traffic measurements make it difficult to obtain comprehensive datasets for analysis and modeling. Generative approaches offer a promising alternative by supplementing scarce real-world measurements, improving simulation fidelity, and facilitating the design and evaluation of network control strategies in a cost-effective and privacy-preserving manner~\cite{10.1145/3419394.3423643}. Such generative modeling supports various downstream optimization tasks, including Base Station (BS) deployment~\cite{10086045, sleeping-1, sleeping-2, sleeping-3}, resource allocation~\cite{10224322, 9725256}, traffic load balancing~\cite{10976420, 10904019}, and energy-efficient scheduling~\cite{11062900}.
\begin{figure}[tb]
\centering
\includegraphics[width=0.85\linewidth]{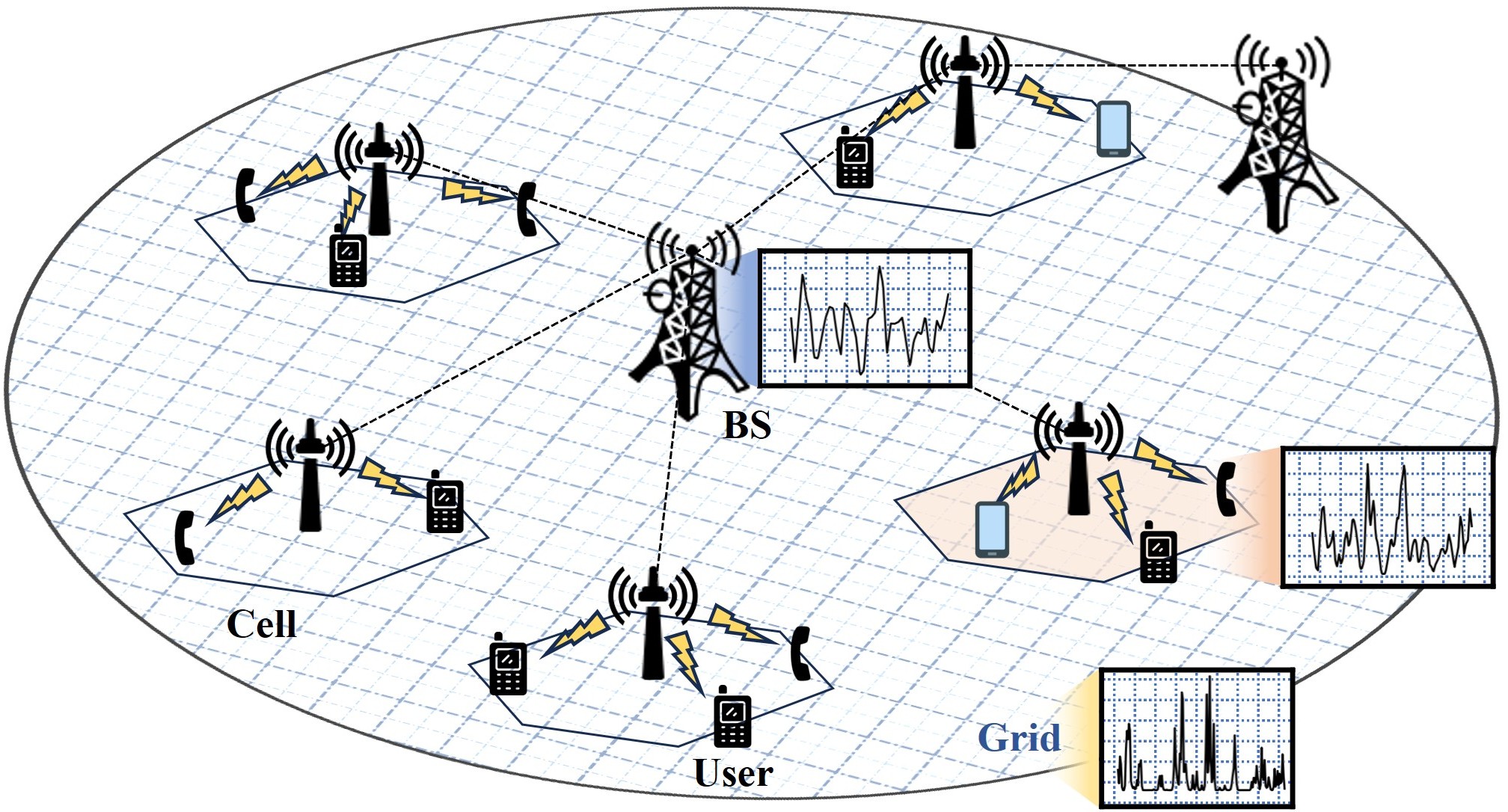}
\caption{Typical cellular mobile network, with multi-layer network nodes including BSs, cells, and users. Grids are aggregations of users within regularly defined geographical areas. The curves in the figure illustrate BS, cell, and grid traffic patterns with different temporal and spatial resolutions.}
\label{Fig_sence}
\vspace{-4mm}
\end{figure}

Over the past decade, extensive efforts have been devoted to generating mobile traffic data across various spatial scales, ranging from city-wide aggregation to regional and BS level granularity~\cite{10.1145/3485983.3494844, 10.1145/3637528.3671544, 10.1145/3703447, 9625773}. Such work typically leverages historical traffic patterns and statistical modeling approaches to simulate spatiotemporal traffic characteristics at diverse granularities. These studies lay foundational groundwork by adequately addressing macroscopic traffic generation demands and demonstrating their value in multiple practical network scenarios~\cite{9844048, 10234717, 10201864}.

However, with the escalating complexity of networks and the emergence of novel mobile services such as immersive XR, autonomous driving, and tactile Internet~\cite{10.1145/3636534.3697422, 10466554, 8584062}, the demand for mobile traffic data generation exhibits two major trends. \textbf{\textit{First, multi-scale traffic generation.}} The typical structure of a cellular mobile network, as shown in Fig.~\ref{Fig_sence}, consists of multiple layers of network nodes, including BSs, cells, grids, and users. As network structures become more complex, joint optimization across multiple network layers can yield greater performance gains compared with traditional single-layer strategies. This trend requires the support of synthetic mobile traffic across multiple network layers. The spatiotemporal resolution differences and inherent correlations among traffic at these layers impose stringent requirements on the cross-layer joint generation and organization of such traffic data. \textbf{\textit{Second, high-resolution traffic generation.}} High-resolution traffic patterns at fine spatial granularities, such as grid-level data, are increasingly critical for precision-based network optimization, including micro-level capacity planning, fine-grained coverage optimization, and user-centric network configuration~\cite{5970249, 10198898}. Yet traffic at refined scales often exhibits intricate spatiotemporal dependencies and distributional complexities, which require generative models to effectively capture and model fine-grained spatial correlations and rapid temporal variations in network dynamics.

Unfortunately, the aforementioned research can only generate coarse-grained mobile traffic with the same single spatiotemporal resolution as the training set. When generating and updating multi-scale traffic data, different models must be used to simulate different network layers, which greatly increases computational overhead. Moreover, generating the complex spatiotemporal distribution of fine-grained traffic exceeds the capabilities of existing methods. It requires advanced user localization techniques~\cite{10.1007/978-3-030-33607-3_45} or specialized network probes~\cite{BOZ2020107158}, incurring substantial communication overhead and computational costs~\cite{9155340, GUERRERO201011}, making it infeasible for continuous collection in large-scale live networks. \textbf{A promising yet largely unexplored alternative is to design a unified model capable of jointly generating multi-scale mobile traffic}, allowing the easily estimated coarse spatiotemporal granularity traffic to progressively guide the generation of finer-grained traffic, thereby achieving multi-scale traffic generation while improving the accuracy of fine-grained traffic synthesis.

\begin{figure}[tb]
\centering
\includegraphics[width=\linewidth]{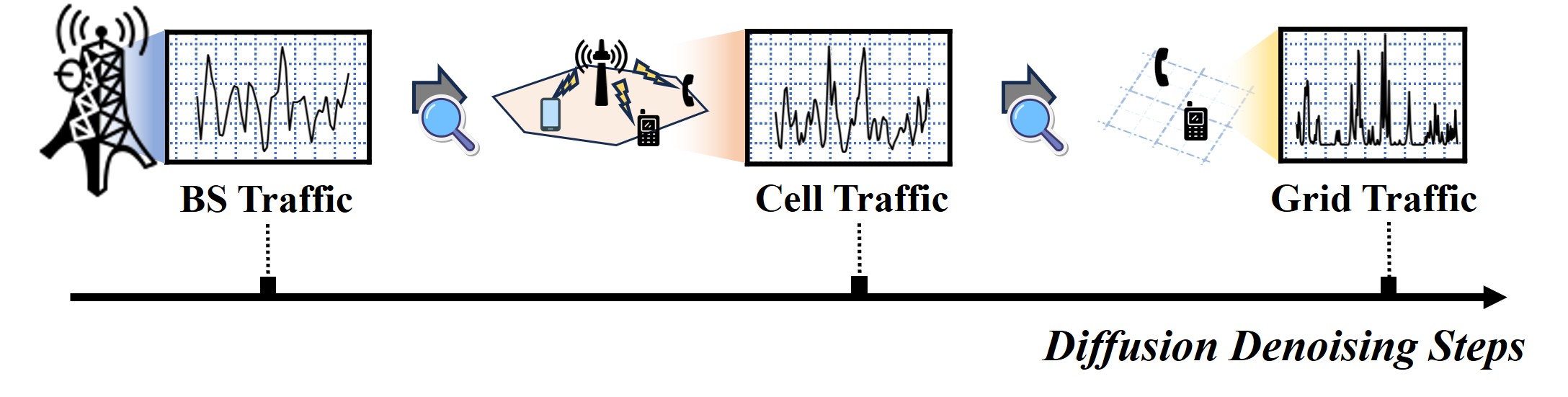}
\caption{Align the spatiotemporal refinement relationships of multi-scale mobile traffic with the denoising process of diffusion models.}
\label{Fig_zoom}
\vspace{-4mm}
\end{figure}

Recently, the emergence of diffusion models has offered promising insights for fine-grained traffic generation. Diffusion models operate by first progressively injecting noise into the original data and subsequently denoising step-by-step to reconstruct high-resolution images from pure noise. This denoising paradigm naturally aligns with the goal of fine-grained mobile traffic synthesis. Inspired by this, we aim to explore a control mechanism for the intermediate states in the noise-adding and denoising processes, as shown in Fig.~\ref{Fig_zoom}, aligning the variation of noise intensity with the change in spatiotemporal resolution of mobile traffic. We seek to make the denoising process progressively \emph{zooms in} on the mobile traffic, revealing increasingly finer spatiotemporal details at each step.

However, achieving this goal presents two main challenges. 
First, it remains unclear how to design effective denoising procedures within the diffusion framework to accurately represent multi-scale mobile traffic. 
Second, it is challenging to model and preserve the cross-resolution dependencies and statistical constraints between traffic at different granularities. 

To address the above challenges, we propose \textbf{ZoomDiff}, a multi-resolution mobile traffic generation model. It achieves a mapping from Urban Environmental Context (UEC) to multi-level mobile traffic, allowing for a \textit{zoom-like} progressive generation for traffic with different scales.
\textbf{First}, we design Denoising Refinement Diffusion Models (DRDM). It is a multi-stage diffusion framework, enabling controllable denoising intermediates through a Hierarchical Noise-adding Process (HNaP) and a Resolution Refinement Denoising Process (RRDP). Multi-scale traffic with different resolutions can be directly extracted from intermediate states at corresponding diffusion steps. 
\textbf{Second}, we design a diffusion prior noise mechanism to establish correlations between adjacent noise-adding and denoising stages. This design enables the coarse-resolution traffic to serve as guidance to control the generation of high-resolution traffic. 
\textbf{Moreover}, we discussed the noise scheduling strategies and the Resolution Greediness Preference of the multi-stage model required for implementing DRDM, and revised the corresponding strategies in conventional diffusion models. We found that the combination of discrete noise intensity, discrete noise-adding, and discrete denoising strategies enables ZoomDiff to achieve optimal multi-scale generation performance. The main contributions can be summarized as follows:

\begin{itemize}
    \item To the best of our knowledge, we propose the first unified model capable of simultaneous generation of multi-scale mobile traffic. Leveraging the sophisticated design of DRDM, it aligns the progressive refinement of spatiotemporal scales of mobile traffic with the denoising process. In other words, multiple intermediate states of the denoising process are constrained to correspond to multi-scale mobile traffic.
    \item We design a prior noise guidance mechanism to establish correlations between noise data at different spatiotemporal resolutions, allowing the high-resolution traffic generation to fully leverage the coarse-grained spatiotemporal patterns learned during the generation of lower-resolution traffic.
    \item Evaluations on real-world datasets demonstrate that the multi-scale traffic generation based on ZoomDiff outperforms state-of-the-art specialized traffic generation models and spatiotemporal generation models. ZoomDiff achieves an average improvement of 18.4\% over the second-best model. 
\end{itemize}

This paper is organized as follows. Section~\ref{sec:rela} provides an overview of related works on mobile traffic generation and spatiotemporal series prediction. Section~\ref{sec:pre} illustrates the preliminaries of our work, including an observation of the traffic correlations, the problem formulation, and an introduction to canonical diffusion models. In Section~\ref{sec:method}, we propose our ZoomDiff models and give detailed technical descriptions of the Environment-aware Conditional Information and the designed DRDM. Section~\ref{sec:eval} shows the evaluation results to verify ZoomDiff's multi-scale generation accuracy and generalizability. Section~\ref{sec:dis} discusses ZoomDiff's capabilities in BS traffic refinement and cross-city generalization performance. Section~\ref{sec:con} provides a conclusion for this paper.

\section{Related Work} 
\label{sec:rela}

\subsection{Mobile Traffic Generation}
\label{sec:rela_1}
Mobile traffic generation aims to learn the environment-aware spatiotemporal dynamics by mining the correlation between mobile traffic and external conditions. Traditional generative algorithms, including mathematical statistics~\cite{10.1145/1159913.1159928, 10.1145/1140086.1140094} and simulation-based approaches~\cite{10.1145/1028788.1028798, 10.1145/1081081.1081129, 10.1145/1231956.1231970}, have limited generalization capabilities. With the development of Machine Learning (ML), generative models such as Generative Adversarial Networks (GANs)~\cite{10.1145/3485983.3494844, 5GT-GAN, Hierarchical-GAN, keGAN} and Transformers~\cite{10946853, UniST} can capture the complex correlations between network demand and environmental features in a data-driven manner. Pandey~\textit{et al.}~\cite{5GT-GAN} and Li~\textit{et al.}~\cite{Hierarchical-GAN} employed GANs to autoregressively model the temporal dynamics of traffic sequences for synthetic traffic generation. Hui~\textit{et al.}~\cite{keGAN} integrated urban environments into Urban Knowledge Graphs (UKGs) for generation guidance. As the stable generative capabilities of diffusion models are validated, recent works have attempted to replace GANs with diffusion-based generators. Zhou~\textit{et al.}~\cite{KSTDiff} applied UKG to the diffusion denoising process for accurate urban population generation. Sivaroopan~\textit{et al.}~\cite{NetDiffus} and Jiang~\textit{et al.}~\cite{NetDiffusion} converted one-dimensional mobile traffic into two-dimensional images to leverage the strong image generation capabilities of diffusion models.

However, existing research primarily focuses on traffic generation at a single spatiotemporal resolution. For the task of generating multi-scale traffic, multiple models or separate generations are typically required. In addition, these models are designed for coarse-scale traffic, such as BS or cell levels, and perform poorly when generating fine-grained traffic at small grid scales. In contrast, ZoomDiff can generate multi-scale mobile traffic in a single process and progressively guide the generation of fine-grained traffic through coarse-scale traffic, achieving both improved generation accuracy and enhanced efficiency.

\subsection{Spatiotemporal Series Prediction/Generation}
\label{sec:rela_mobile_gen}
Mobile traffic generation can be regarded as a controllable spatiotemporal generation task. Numerous studies have employed models such as Convolutional Neural Network (CNN)~\cite{CNN-1, CNN-2, CNN-3}, Recurrent Neural Network (RNN)~\cite{RNN-1, RNN-2, RNN-3}, and Transformers~\cite{Tran-1, Trans-2, Trans-3, Trans-4} for spatiotemporal prediction and generation. Based on GANs, Yoon~\textit{et al.}~\cite{TimeGAN} proposed TimeGAN, which combines the flexibility of unsupervised paradigms with the control of supervised training by jointly optimizing supervision and adversarial learning in the embedding space. Saxena~\textit{et al.}~\cite{2019arXiv190708556S} introduced D-GAN, which jointly learns data generation and variational inference. In recent years, an increasing number of works have explored diffusion models to capture spatiotemporal distribution patterns. Yuan~\textit{et al.}~\cite{10.1145/3580305.3599511} improved the spatiotemporal point process (STPP) using diffusion models to learn complex joint spatiotemporal distributions in urban data. Yuan~\textit{et al.}~\cite{Diffusion-TS} reconstructed samples at each diffusion step and introduced Fourier-based loss terms to capture temporal dynamics. In addition to direct modeling in the temporal and spatial domains, learning spatiotemporal patterns through state space is becoming increasingly popular. Inspired by deep state space models (S4), Zhou~\textit{et al.}~\cite{10.5555/3618408.3620205} proposed LS4, which enhances spatiotemporal modeling capabilities via latex
Ordinary Differential Equation (ODE) evolution in state space. 

Researchers have attempted to achieve multi-resolution prediction of time series. Shen~\textit{et al.}~\cite{ICLR2024_d64740dd} utilized cascaded diffusion models that progressively refine the temporal resolution of the series. Shabani~\textit{et al.}~\cite{2022arXiv220604038S} proposed Scaleformer, in which a set of Transformer modules is used to iteratively refine time series across different temporal scales. However, these cascaded architectures linearly increase the number of model parameters as the resolution hierarchy grows and are prone to error accumulation. Fan~\textit{et al.}~\cite{2024arXiv240305751F} proposed MG-TSD, which leverages the inherent temporal granularity levels within the data as predefined targets at intermediate diffusion steps. Nevertheless, this model focuses solely on generating time-domain multi-resolution sequences without incorporating spatial multi-scale information. Li~\textit{et al.}~\cite{MDPNet} introduced MDPNet, which leverages spatial hierarchy to guide an ODE system in learning spatial semantics and performs forecasting using a diffusion-based encoder. However, the multi-scale spatial information in this approach only serves as auxiliary input for understanding spatial distributions, and the model does not produce multi-resolution outputs.

\section{PRELIMINARY} 
\label{sec:pre}
In this section, we first qualitatively validate the spatiotemporal correlations across different scales of mobile traffic. Then, the formulation of the canonical diffusion models is briefly outlined. Finally, we present a problem definition for multi-scale mobile traffic generation.

\subsection{Correlations among Multi-scale mobile traffic}
\begin{figure}[tb]
\centering
\includegraphics[width=0.9\linewidth]{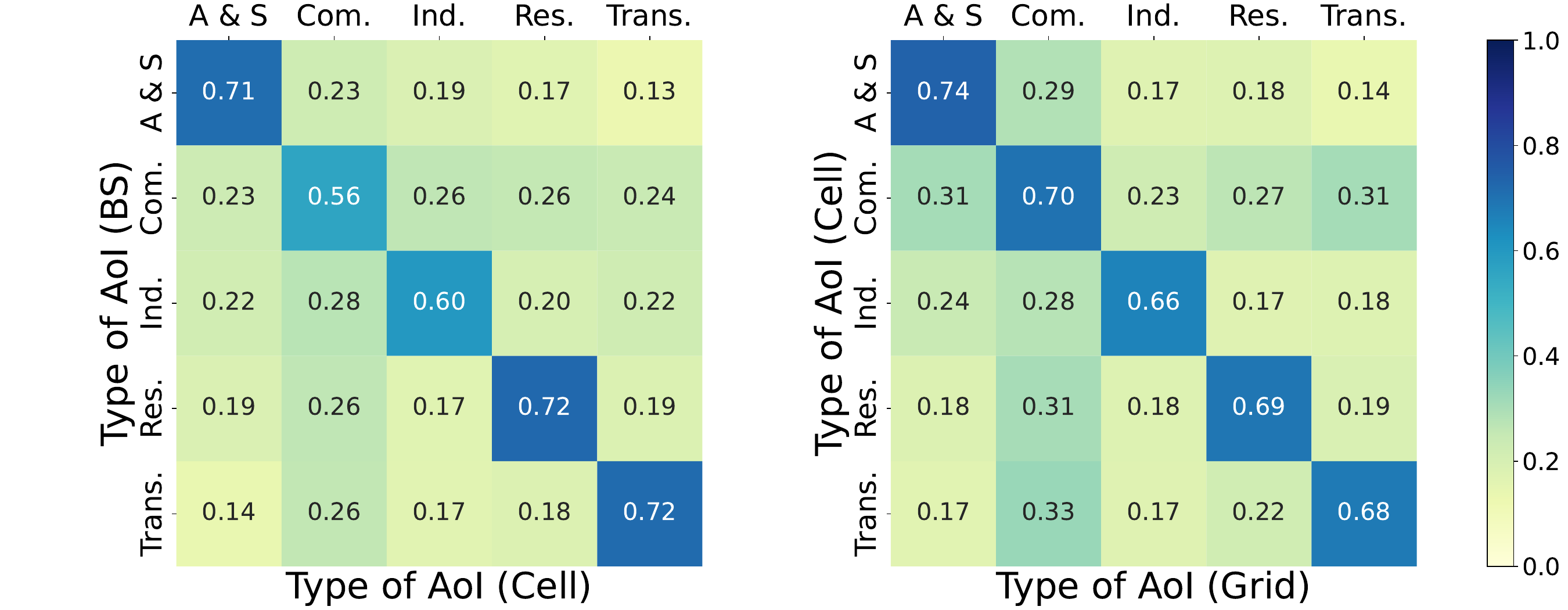}
\caption{Heatmaps of RV-coefficients between different levels of traffic across various AoI types.}
\label{Fig_heatmap}
\vspace{-4mm}
\end{figure}

\begin{figure}[tb]
\centering
\includegraphics[width=0.9\linewidth]{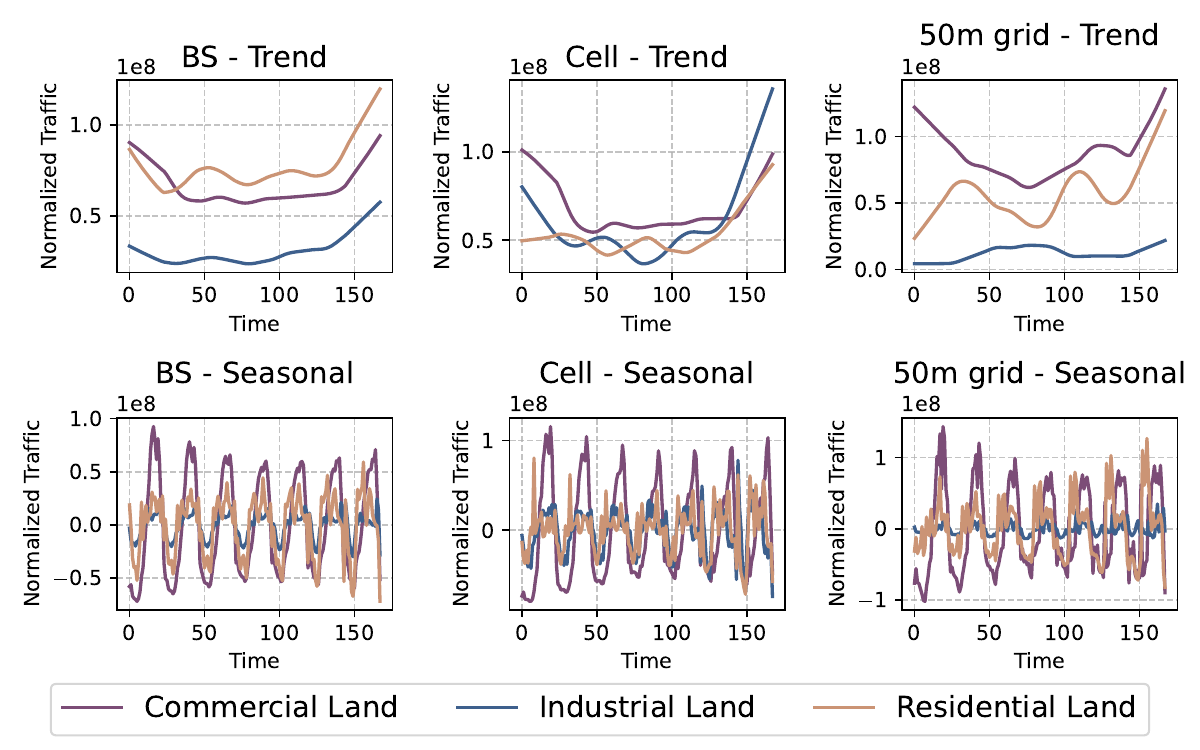}
\caption{Time series decomposition of BS, Cell, and 50m grid traffic.}
\label{Fig_seasonal}
\vspace{-4mm}
\end{figure}

The mobile traffic across different scales, including the BS traffic, cell traffic, and grid-level traffic, exhibits a top-down aggregation relationship due to the inherent hierarchy among their service areas. Figure~\ref{Fig_seasonal} shows the temporal decomposition of hourly BS traffic, cell traffic, and 50m grid traffic from three kinds of Area of Interest (AoI) over one week. The trend component reflects the distribution of daily average traffic over seven days, while the seasonal component captures the temporal patterns of daily traffic after removing the trend. To be quantitative, Fig.~\ref{Fig_heatmap} computes the RV-coefficients~\cite{10.2307/2347233} between adjacent levels of mobile traffic. It can be observed that traffic at neighboring scales within the same AOI exhibits high similarity. This provides the theoretical foundation for jointly generating multi-scale traffic, where coarse-grained traffic can guide the generation of fine-grained traffic. The RV-coefficient is defined as:
\begin{equation}
    \small
    \mathcal{R}(\Gamma_1, \Gamma_2) = \frac{\text{tr}(\Gamma_1 \Gamma_1^\intercal \Gamma_2 \Gamma_2^\intercal)}{\sqrt{\text{tr}((\Gamma_1 \Gamma_1^\intercal)^2) \cdot \text{tr}((\Gamma_2 \Gamma_2^\intercal)^2)}},
\end{equation}
which is used to measure the similarity between multidimensional data matrices. $\Gamma_1$ and $\Gamma_2$ denote two high-dimensional data matrices, and $\text{tr}$ computes the trace of a matrix. 

\subsection{Problem Formulation}
Mobile traffic at different layers of network nodes reflects user demand distributions at different spatiotemporal scales. Given the UEC $\pmb{c}$, the objective of the model is to learn a mapping from UEC to the corresponding regional mobile traffic $\pmb{x}$, and simultaneously generate multi-level traffic within the region at different spatiotemporal resolutions. Consider a mobile network with $K \geq 2$ levels, let $\pmb{x}_{s_{\delta_k} t_{\tau_k}}$ denote the mobile traffic at level $k$ with spatial resolution $s_{\delta_k}$ and temporal resolution $t_{\tau_k}$, where $1 \leq k \leq K$, $\delta_k \geq \delta_{k-1}$ and $\tau_k \geq \tau_{k-1}$. This mapping can be expressed as $\pmb{c} \mapsto (\pmb{x}_{s_{\delta_1} t_{\tau_1}}, ..., \pmb{x}_{s_{\delta_{K-1}} t_{\tau_{K-1}}}, \pmb{x}_{s_{\delta_K} t_{\tau_K}})^{T \times H \times W}$, where $T$ denotes the temporal length of the finest traffic series, and $H$, $W$ represent the finest spatial dimensions. As the spatiotemporal resolution increases, the traffic becomes increasingly influenced by fine-grained urban environments and user behavioral characteristics. Therefore, generating high-dimensional spatiotemporal traffic at fine resolutions is a complex task.

This multi-level traffic generation problem can be abstracted as a conditional generation problem of multi-resolution spatiotemporal sequences. Given the environmental condition $\pmb{c}$, the $K$-stage multi-resolution generation problem can be formulated as:
\begin{equation}
\small
    (\pmb{x}_{s_{\delta_1} t_{\tau_1}}, ..., \pmb{x}_{s_{\delta_{K-1}} t_{\tau_{K-1}}}, \pmb{x}_{s_{\delta_K} t_{\tau_K}})=\mathcal{\pmb{\Psi}}(\pmb{c}, K),
\end{equation}
where $\mathcal{\pmb{\Psi}}$ denotes the generative model that performs unified modeling and generation of traffic data across all resolutions.

\subsection{Canonical Diffusion Models}
\label{sec::SRDM}
The core idea of canonical diffusion models~\cite{DDPM, CSDI, DiT} is to construct a forward diffusion process that gradually perturbs the original data into Gaussian noise, while simultaneously training a learnable reverse denoising process to recover the original data. According to Ho~\textit{et al.}~\cite{DDPM}, the forward diffusion process can be obtained as:
\begin{equation}
\small
    \pmb{x}^{(n)} = \sqrt{\alpha^{(n)}} \pmb{x}^{(0)} + \sqrt{1 - \alpha^{(n)}} \boldsymbol{\epsilon}, \quad \boldsymbol{\epsilon} \sim \mathcal{N}(0, \pmb{I}),
    \label{eq::forward}
\end{equation}
where $n$ is the order of diffusion steps, $\pmb{x}^{(n)}$ is the noised sample at step $n$, $\alpha^{(n)} = \prod_{i=1}^n \hat{\alpha}_i$ is a predefined noise scheduling coefficient that controls the noise intensity at each step. After a sufficient number of steps $N$, the data distribution converges toward a standard normal distribution $\pmb{x}^{(N-1)} =\boldsymbol{\epsilon}\sim \mathcal{N}(0, \pmb{I})$.

During the forward process, the diffusion model learns a parameterized neural network $\boldsymbol{\epsilon}_\theta(\pmb{x}^{(n)}, n)$ to approximate the noise added in each step. $\theta$ means the model parameters. The training objective is typically formulated as a mean-squared error loss for noise reconstruction:
\begin{equation}
\small
    \mathcal{L}_{\text{s}} = \mathbb{E}_{\pmb{x}^{(0)}, \boldsymbol{\epsilon}, n} \left[ \left\| \boldsymbol{\epsilon}_\theta(\pmb{x}^{(n)}, n) - \boldsymbol{\epsilon} \right\|_2^2 \right].
\end{equation}

In the reverse process, the model denoises from a standard Gaussian noise $\pmb{x}^{(N-1)} =\boldsymbol{\epsilon}$ and performs iterative reverse sampling according to~\cite{DDPM}:
\begin{equation}
\small
    \pmb{x}^{(n-1)} = \frac{1}{\sqrt{\hat\alpha^{(n)}}} \left( \pmb{x}^{(n)} - \frac{1 - \hat\alpha^{(n)}}{\sqrt{1 - \alpha^{(n)}}} \boldsymbol{\epsilon}_\theta(\pmb{x}^{(n)}, n) \right) + \sigma^{(n)} \boldsymbol{\epsilon},
    \label{eq::reverse}
\end{equation}
where $\sigma^{(n)}=\sqrt{1-\hat\alpha^{(n-1)}}(1-\alpha^{(n-1)})/({1-\alpha^{(n)}})$. 

\emph{Limitation}.
Canonical diffusion models are fundamentally designed for \textit{single-resolution} data generation. That is, they progressively denoise random noise at a fixed resolution throughout the diffusion process. However, the intermediate states during denoising are typically unstructured and semantically meaningless, offering no interpretable representation of coarser or finer information. Moreover, the models lack mechanisms to model dependencies across different spatiotemporal resolutions, treating each step as an isolated transformation. As a result, canonical diffusion frameworks are inherently limited in addressing the requirements of the multi-resolution mobile traffic generation problem.

\subsection{Technical Challenges}
To address the problem of multi-scale mobile traffic generation, it is necessary to design a novel diffusion model architecture that goes beyond the limitations of canonical frameworks. However, this design process still faces the following three key challenges:

\begin{itemize}
    \item How to leverage diffusion models to construct a generative framework for multi-resolution spatiotemporal data;
    \item How to enable the model to understand the intrinsic relationships among multi-resolution data;
    \item How to equip the model with the ability to capture spatiotemporal dependencies in high-dimensional traffic sequences.
\end{itemize}

\section{Method} 
\label{sec:method}
\begin{figure}[tb]
\centering
\includegraphics[width=0.85\linewidth]{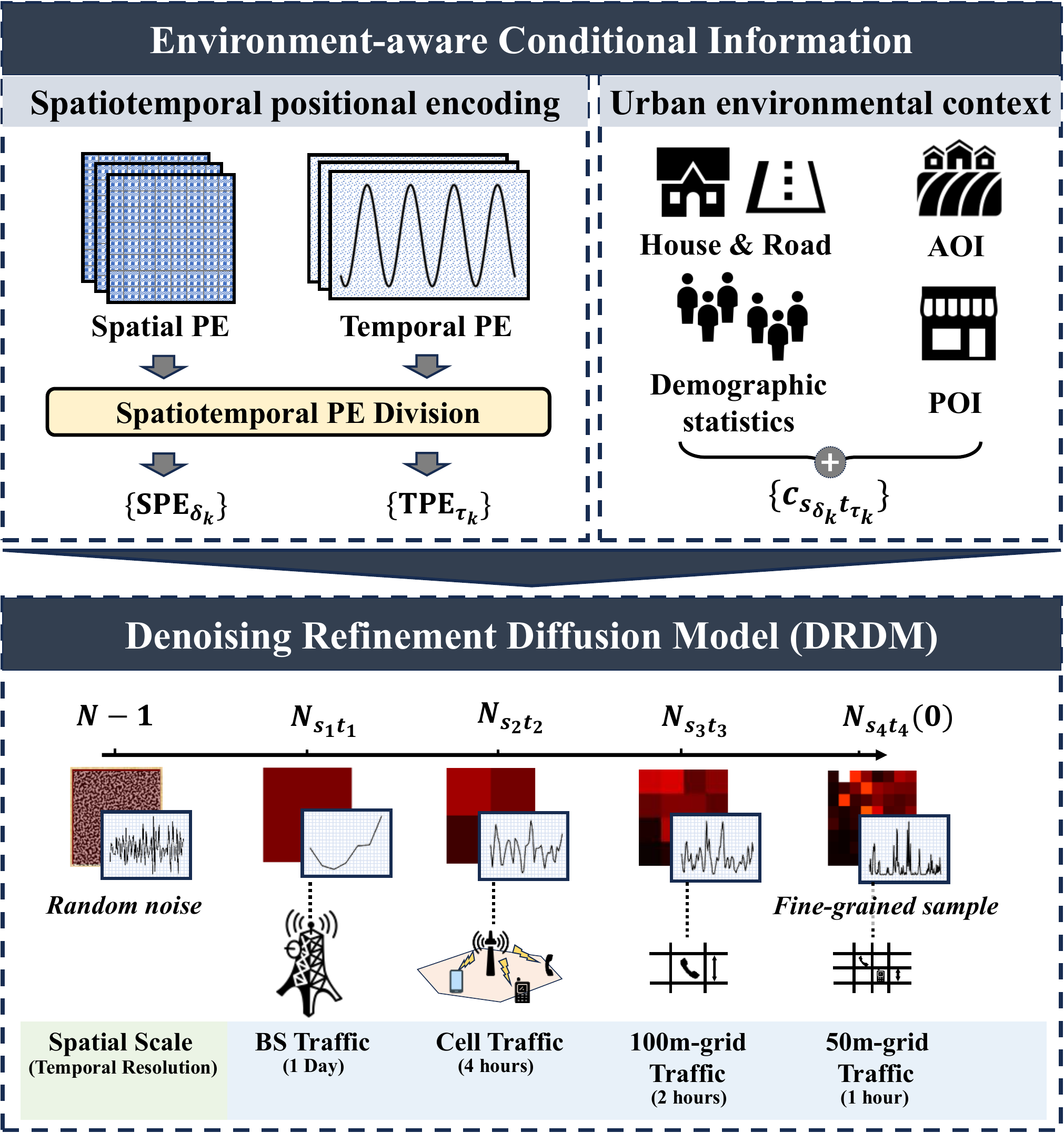}
\caption{Overall framework of ZoomDiff. It consists of two components: the Environment-aware Conditional Information and the Denoising Refinement Diffusion Models (DRDM).}
\label{Fig_zoomdiff}
\vspace{-4mm}
\end{figure}

To address the technical challenges, we propose ZoomDiff, a diffusion-based model for multi-resolution mobile traffic generation. ZoomDiff restructures the intermediate states of the denoising process so that they are no longer disordered noisy data, but rather a series of data representations with progressively increasing resolution throughout the denoising trajectory. In other words, it introduces a customized denoising process that gradually zooms in on finer details of regional mobile traffic. Fig.~\ref{Fig_zoomdiff} shows the overall framework of ZoomDiff. It consists of two components: the Environment-aware Conditional Information and the DRDM. DRDM takes a series of conditions that are aligned with the spatiotemporal resolutions of the network scales as input. It outputs multi-scale mobile traffic with various spatiotemporal resolutions through different intermediate denoising states, and ultimately produces a fine-grained sample corresponding to the smallest spatial scale and the finest temporal resolution.

\subsection{Environment-aware Conditional Information}
The environment-aware conditional information consists of the Spatiotemporal Positional Encoding (StPE) and the UEC. The StPE is designed to enable the diffusion model to capture temporal characteristics and spatial relationships in high-dimensional traffic data. The UEC is derived from multidimensional urban data, which is relevant to mobile users' traffic demands.

\subsubsection{Spatiotemporal Positional Encoding (StPE)}
StPE is composed of a Temporal Positional Encoding (TPE) and a Spatial Positional Encoding (SPE). The TPE is constructed using a 128-D sinusoidal encoding, which is formulated as 
\begin{equation}
\small
    \text{TPE}(t)= \left[ \sin\left(t/10000^{\frac{2i}{128}} \right),\ \cos\left(t/10000^{\frac{2i}{128}} \right) \right]_{i=0}^{63},
\end{equation}
and the SPE is implemented as a 128-D learnable embedding, which can be represented as $\text{SPE} \in \mathbb{R}^{128 \times H \times W}$. Different SPEs are used for different spatial resolutions. TPE and SPE are concatenated to form a high-dimensional StPE: $[\text{TPE}, \text{SPE}]^{256\times L \times H \times W}$. This encoding is further coarsened via a downsampling-based Spatiotemporal PE Division to align with the multi-resolution traffic, yielding encoding series $\{\text{TPE}_{\tau_k}\}$ and $\{\text{SPE}_{\delta_k}\}$. They help ZoomDiff to understand various spatiotemporal relations across resolutions. The formulations of TPE and SPE are given as follows.

\subsubsection{Urban Environment Context (UEC)}
Considering generality and accessibility, we construct the UEC using the following four types of urban data: House \& Road type $c_{S} \in \mathbb{N}^{H \times W}$, AOI types $c_{\text{AoI}}\in \mathbb{N}^{H \times W}$, POI embedding $c_\text{PoI}\in \mathbb{N}^{d \times H \times W}$, and historical demographic statistics $c_{\text{pop}}\in \mathbb{N}^{T \times H \times W}$. $c_{S} \in \mathbb{N}^{H \times W}$ indicates information about indoor and outdoor areas as well as road types; $c_{\text{AoI}}\in \mathbb{N}^{H \times W}$ represent the functional zoning of different urban areas; $c_\text{PoI}\in \mathbb{N}^{d \times H \times W}$ reflects the types and quantity distribution of urban entities surrounding a region. $c_{\text{pop}}\in \mathbb{N}^{T \times H \times W}$ represents the historical statistical distribution of the regional population. To obtain a unified representation, we use a set of MLP-based mappers $\{f_i\}_1^4$ to project the above data into the same latent space. They are then combined to obtain high-resolution urban environment semantics: $\pmb{c}=f_1(c_{S}) \oplus f_2(c_{\text{AoI}}) \oplus f_3(c_\text{PoI}) \oplus f_4(c_{\text{pop}})$. We further decompose $\pmb{c}$ into a series of UEC $\{\pmb{c}_{s_{\delta_k} t_{\tau_k}}\}$ aligned with the target spatiotemporal resolutions through downsampling. It could guide the conditional distribution fitting at each stage of the DRDM.

\subsection{Denoising Refinement Diffusion Models (DRDM)}
\begin{figure*}[tb]
\centering
\includegraphics[width=0.8\linewidth]{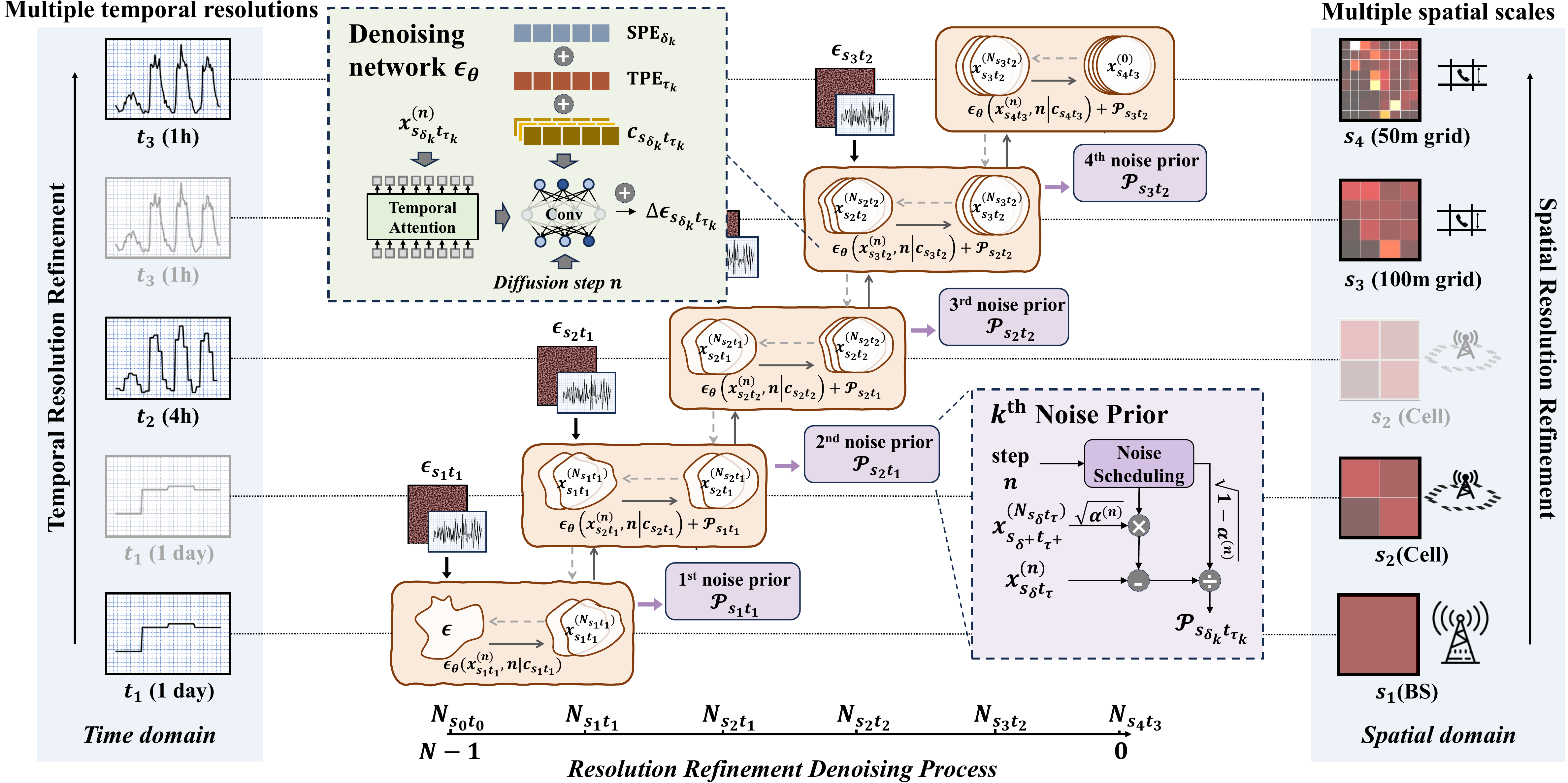}
\caption{The pipeline of the multi-stage denoising process of DRDM. Different stages correspond to different spatial scales and temporal resolutions, where the gray color indicates that the spatial scale or temporal resolution at that stage remains unchanged compared to the previous stage.}
\label{Fig_DRDM}
\vspace{-4mm}
\end{figure*}

DRDM is a multi-resolution diffusion model architecture. The denoising process of DRDM is shown in Fig.~\ref{Fig_DRDM}. It restructures the intermediate states of the denoising process, allowing a single denoising trajectory to generate samples with varying spatiotemporal resolutions at different nodes. The multi-resolution data generation capability of DRDM is achieved through a customized multi-stage forward and reverse process, including the \textit{HNaP} and the \textit{RRDP}. DRDM uses a customized denoising network to estimate noise from multi-dimensional conditions. It further incorporates a noise prior guidance mechanism to establish connections across stages with different spatiotemporal resolutions. This section provides a detailed introduction to the noise prior guidance mechanism, HNaP, and RRDP, and discusses specialized noise scheduling and Resolution Greediness Preference strategies adapted for DRDM.

\subsubsection{Diffusion Noise Prior Guidance}
In Section~\ref{sec::SRDM}, Eq.~\ref{eq::forward} presents the general form of the forward process in diffusion models. Inspired by~\cite{2025arXiv250113794S}, suppose the real data $\pmb{x}^{(0)}$ contains a fixed, prior-known component $\pmb{H}$, such that $\pmb{x}^{(0)} = \pmb{H} + \Delta\pmb{x}^{(0)}$, where $\Delta\pmb{x}^{(0)}$ denotes the residual between $\pmb{x}^{(0)}$ and $\pmb{H}$. Then, Eq.~\ref{eq::forward} can be rewritten as:
\begin{equation}
\small
    \pmb{x}^{(n)} 
    = \sqrt{\alpha^{(n)}} 
    \underbrace{(\pmb{H} + \Delta\pmb{x}^{(0)})}_{\pmb{x}^{(0)}}
    + \sqrt{1 - \alpha^{(n)}} \boldsymbol{\epsilon}.
    \label{eq::forward_DRDM}
\end{equation}
By equivalently transforming the above equation, we obtain:
\begin{equation}
\small
    \begin{aligned}
    \boldsymbol{\epsilon} =& \frac{\pmb{x}^{(n)} - \sqrt{\alpha^{(n)}} (\pmb{H} + \Delta\pmb{x}^{(0)})}{\sqrt{1 - \alpha^{(n)}}} \\
    =& 
    \underbrace{
        \frac{\pmb{x}^{(n)} - \sqrt{\alpha^{(n)}} \Delta\pmb{x}^{(0)}}{\sqrt{1 - \alpha^{(n)}}}
    }_{\Delta\boldsymbol{\epsilon}}
    -
    \underbrace{
        \frac{\sqrt{\alpha^{(n)}} \pmb{H}}{\sqrt{1 - \alpha^{(n)}}}
    }_{\hat{\boldsymbol{\epsilon}}}
    \\
    :=& \Delta\boldsymbol{\epsilon} - \hat{\boldsymbol{\epsilon}},
    \end{aligned}
    \label{eq::epsilon_DRDM}
\end{equation}

where $\hat{\epsilon}$ is defined as the prior noise component associated with the fixed part $\pmb{H}$. The model's learning objective is to fit the residual noise $\Delta\pmb\epsilon = \pmb\epsilon + \hat{\pmb\epsilon}$. The prior noise mechanism introduces prior knowledge into both the noise-adding and denoising processes, enabling the diffusion model to generate data biased toward a given prior distribution. The loss function of the diffusion model with the applied prior noise mechanism in the forward process can be formulated as:
\begin{equation}
\small
    \mathcal{L}_{\text{m}} = \mathbb{E}_{\pmb{x}^{(0)}, \boldsymbol{\epsilon}, n} \left[ \left\| \boldsymbol{\epsilon}_\theta(\pmb{x}^{(n)}, n) - (\boldsymbol{\epsilon} + \boldsymbol{\hat{\epsilon}}) \right\|_2^2 \right],
    \label{eq_loss_prior}
\end{equation}
where the objective of the denoising network $\boldsymbol{\epsilon}_\theta(\pmb{x}^{(n)}, n)$ is to approximate the noise residual $\Delta\pmb\epsilon$.

\subsubsection{Hierarchical Noise-adding Process}
\label{sec::zoomdiff_forward}
To enable the reduction of intermediate states in the denoising process, we first introduce a customized design for the noise-adding process in DRDM. Given a $K$-level ordered sequence tuple with progressively refined resolutions $\{\pmb{x}\}_K=\{\pmb{x}_{s_1 t_1}, ..., \pmb{x}_{s_{\delta_{K-1}} t_{\tau_{K-1}}}, \pmb{x}_{s_{\delta_K} t_{\tau_K}}\}$, where $\pmb{x}_{s_{\delta_K} t_{\tau_K}}$ has the highest spatiotemporal resolution and serves as the starting point of noise addition. Unlike the forward process in canonical diffusion models, HNaP is a multi-stage diffusion process. The total number of diffusion steps $N$ is divided into multiple segments corresponding to the resolution levels in $\{\pmb{x}\}_K$. These segment boundaries are denoted by $\{N\}_K=\{N_{s_{\delta_K} t_{\tau_K}}(=0),N_{s_{\delta_{K-1}} t_{\tau_{K-1}}}, ..., N_{s_1 t_1}\}$, with $N_{s_0 t_0}=N$.

During the $k$-th noise-adding sub-process in HNaP, where $k \in \{1, 2, ..., K\}$, the diffusion step $n \in [N_{s_{\delta_k} t_{\tau_k}}, N_{s_{\delta_{k-1}} t_{\tau_{k-1}}})$. This sub-process takes $\pmb{x}_{s_{\delta_K} t_{\tau_K}}$ as the initial state for noise addition. After iterating through all $k$, HNaP constructs a trajectory starting from $\pmb{x}_{s_{\delta_K} t_{\tau_K}}$ and ending at $\boldsymbol{\epsilon} \sim \mathcal{N}(0, \pmb{I})$, with intermediate states representing data at various spatiotemporal resolutions. To associate successive stages, we adopt a diffusion prior mechanism. In the $k$-th noise-adding sub-process, we treat $\pmb{x}_{s_{\delta_{k+1}} t_{\tau_{k+1}}}$ as a prior bias and define the $k^{\text{th}}$ prior noise as
\begin{equation}
\small
    \mathcal{P}_{s_{\delta_{k-1}} t_{\tau_{k-1}}} = 
    \begin{cases}
    \frac{\sqrt{\alpha^{(n)}_{k}} \pmb{x}_{s_{\delta_{k-1}} t_{\tau_{k-1}}}}{\sqrt{1 - \alpha^{(n)}_{k}}}, & \text{if } k > 1 \\
    0, & \text{if } k = 1
    \end{cases}.
\end{equation}
Then, HNaP can be formally formulated as
\begin{equation}
\small
    \pmb{x}_{s_{\delta_k} t_{\tau_k}}^{(n)} = \sqrt{\alpha^{(n)}_k} \pmb{x}^{(0)}_{s_{\delta_k} t_{\tau_k}} + \sqrt{1 - \alpha^{(n)}_k} (\pmb\epsilon+\mathcal{P}_{s_{\delta_{k-1}} t_{\tau_{k-1}}}).
    \label{eq::HNaP}
\end{equation}
Under this formulation, the training objective becomes a diffusion-prior-based loss:
\begin{equation}
    \small
    \begin{aligned}
        &\mathcal{L}_{\text{HNap}} = \\
        &\mathbb{E}_{\pmb{x}_{s_{\delta_k} t_{\tau_k}}^{(0)}, \boldsymbol{\epsilon}, n} \left[ \left\| (\boldsymbol{\epsilon}_\theta(\pmb{x}_{s_{\delta_k} t_{\tau_k}}^{(n)}, n|\pmb{c}_{s_{\delta_k} t_{\tau_k}})+\mathcal{P}_{s_{\delta_{k-1}} t_{\tau_{k-1}}}) - \boldsymbol{\epsilon} \right\|_2^2 \right].
    \end{aligned}
    \label{eq::loss_HNaP}
\end{equation}
In HNaP, the step segmentation method $\{N\}_K$ is a mapping over $k$, which we refer to as Resolution Greediness Preference (RGP) in Multi-stage Denoising; similarly, the noise coefficient $\alpha^{(n)}_k$ is also a mapping over $k$, referred to as Noise Scheduling. In Section~\ref{sec::ns} and Section~\ref{sec::rgp}, we provide a detailed discussion and analysis of different noise scheduling strategies and diffusion step distributions.

\subsubsection{Resolution Refinement Denoising Process}
\label{sec::zoomdiff_reverse}
HNaP provides $K-1$ controllable intermediate states with progressively coarsened spatiotemporal resolutions for the forward process of DRDM. Since the noise-adding and denoising processes in diffusion models are symmetric, the reverse process of DRDM also contains $K-1$ controllable intermediate states, but with gradually refined spatiotemporal resolutions. We refer to this process as the RRDP. Similar to Eq.~\ref{eq::reverse}, RRDP can be formally expressed as:
\begin{equation}
\small
    \pmb{x}^{(n-1)}_{s_{\delta_k} t_{\tau_k}} = \frac{1}{\sqrt{\alpha^{(n)}_k}} \left(\pmb{x}^{(n)}_{s_{\delta_k} t_{\tau_k}} - \frac{1 - \alpha^{(n)}_k}{\sqrt{1 - \bar{\alpha}^{(n)}_k}} \Sigma_{k}\right) + \sigma^{(n)}_k \boldsymbol{\epsilon},
    \label{eq::reverse_DRDM}
\end{equation}
where $\boldsymbol{\epsilon} \sim \mathcal{N}(0, \pmb{I})$, $\Sigma_{k}=\boldsymbol{\epsilon}_\theta(\pmb{x}^{(n)}_{s_{\delta_k} t_{\tau_k}}, n)+ \mathcal{P}_{s_{\delta_{k-1}} t_{\tau_{k-1}}}$, and $\sigma^{(n)}_k=\sqrt{1-\hat\alpha^{(n-1)}_k}(1-\alpha^{(n-1)}_k)/({1-\alpha^{(n)}_k})$. During the execution of the reverse process, $n$ is iterated backward from $N-1$ to $0$, while $k$ increases step by step from $1$ to $K$ according to the value of $n$. The prior noise $\mathcal{P}_{s_{\delta_k} t_{\tau_k}}$ can be computed from the denoised sample at the adjacent stage $\pmb{x}_{s_{\delta_k} t_{\tau_k}}^{N_{s_{\delta_k} t_{\tau_k}}}$. Based on the above process, RRDP allows us to directly extract $K$ denoised outputs with different spatiotemporal resolutions at steps $n=N_{s_1 t_1}, ..., N_{s_{\delta_{K-1}} t_{\tau_{K-1}}}, N_{s_{\delta_K} t_{\tau_K}}(0)$.

\subsubsection{Denoising Network}
The denoising network is the core component of a diffusion model responsible for progressively restoring the data distribution structure from noisy samples. As shown in Fig.~\ref{Fig_DRDM}, DRDM shares the same denoising network $\boldsymbol{\epsilon}_\theta$ across all stages, while controlling it with inputs of different spatiotemporal scales to estimate noise at each stage. It maps the noisy data $\pmb{x}_{s_{\delta_k} t_{\tau_k}}$, the UEC (including $\text{TPE}_{\tau_k}$, $\text{SPE}_{\delta_k}$, and $\pmb{c}_{s_{\delta_k} t_{\tau_k}}$), and the diffusion step $n$ into a unified embedding space through convolutional layers. These embeddings are then fused by element-wise addition. The convolutional layers also act as noise fitters, mapping the fused embeddings at the output layer to the estimated noise residual $\Delta\pmb\epsilon_{s_{\delta_k} t_{\tau_k}}$. The inclusion of conditional terms enables the denoising network to estimate the conditional noise distribution under specific external conditions $\pmb{c}_{s_{\delta_k} t_{\tau_k}}$, expressed as $\boldsymbol{\epsilon}_\theta(\pmb{x}_{s_{\delta_k} t_{\tau_k}}^{(n)}, n|\pmb{c}_{s_{\delta_k} t_{\tau_k}})$. We add $\Delta\pmb\epsilon_{s_{\delta_k} t_{\tau_k}}$ to the prior noise $\mathcal{P}_{s_{\delta_{k-1}} t_{\tau_{k-1}}}$ and train the denoising network within HNaP using Eq.~(\ref{eq::loss_HNaP}). RRDP employs the trained denoising network to recover the original data from random noise through Eq.~(\ref{eq::reverse_DRDM}).

\subsection{Noise Scheduling for DRDM}
\label{sec::ns}
\begin{figure}[tb]
\centering
\includegraphics[width=\linewidth]{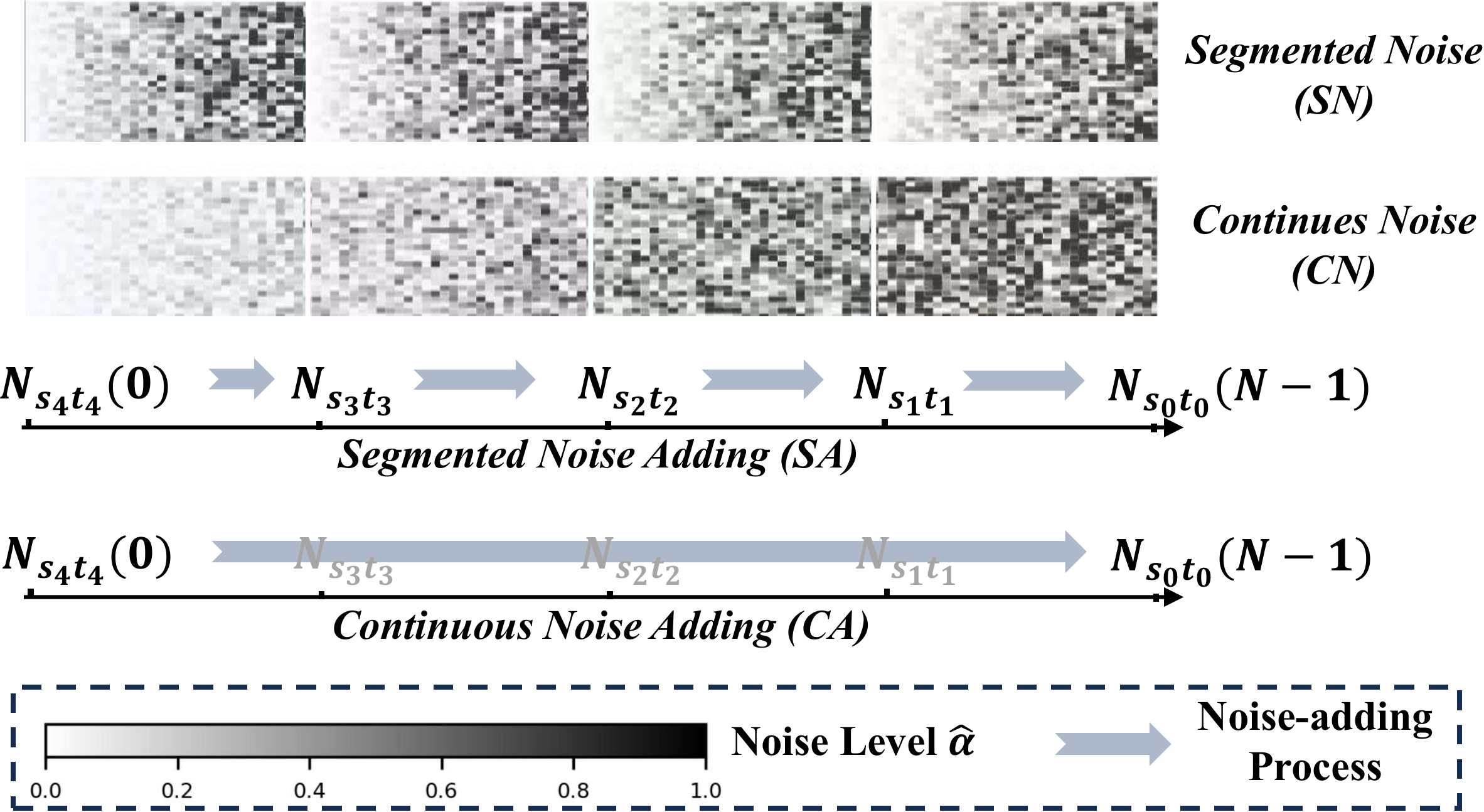}
\caption{Noise scheduling for DRDM in the noise-adding process.}
\label{Fig_nas}
\end{figure}

\begin{figure}[tb]
\centering
\includegraphics[width=\linewidth]{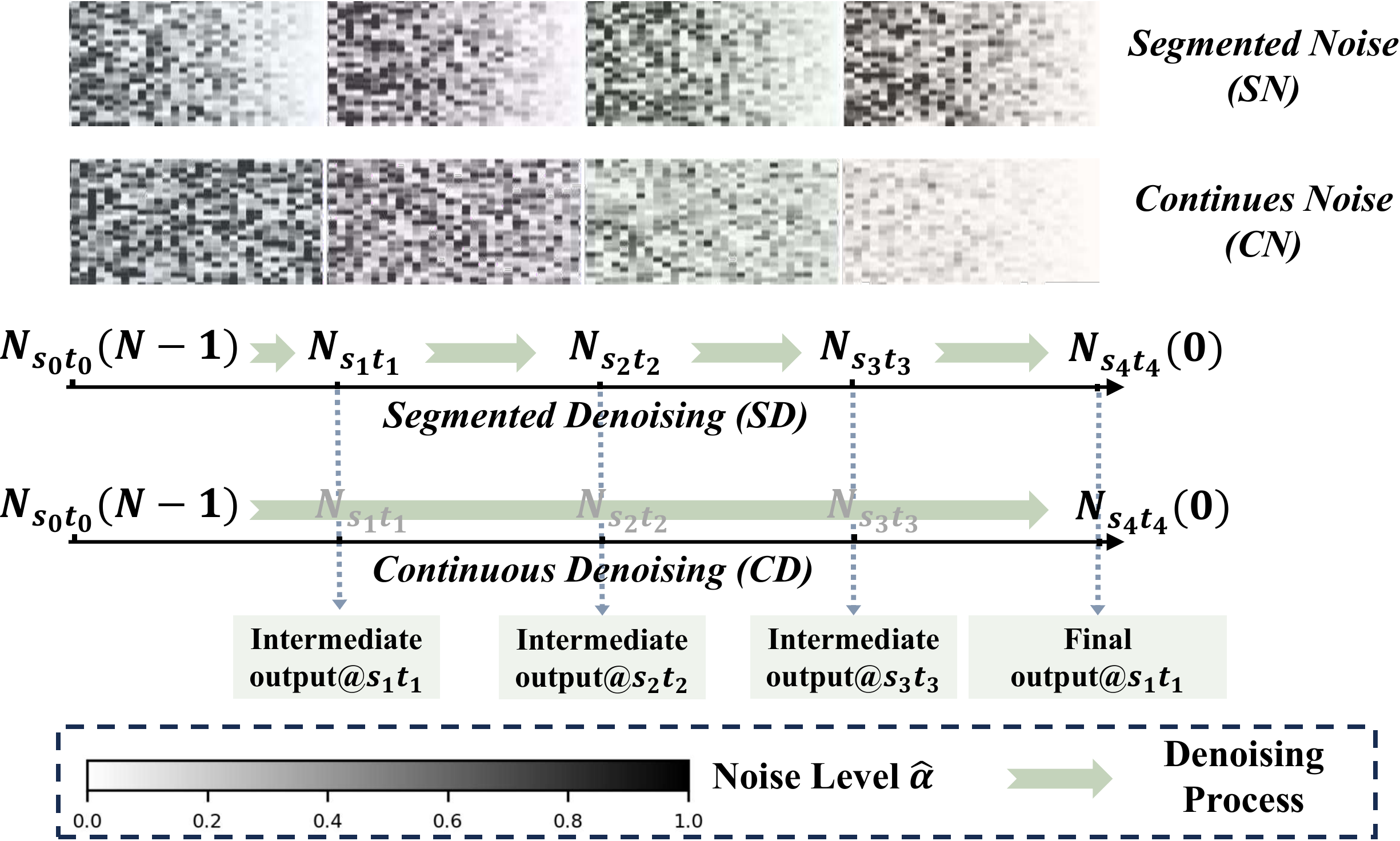}
\caption{Noise scheduling for DRDM in the denoising process.}
\label{Fig_ds}
\vspace{-4mm}
\end{figure}

Noise scheduling is a mechanism in diffusion models that defines how noise intensity varies over time between the forward and reverse processes. Canonical diffusion models typically adopt a continuous linear schedule, where the noise strength $\hat\alpha \in (0, 1)$ increases linearly with $n$, and the data is progressively and continuously noised according to Eq.~\ref{eq::forward}. However, since DRDM segments the denoising process, traditional noise scheduling strategies may no longer be suitable. To effectively constrain the intermediate states of the denoising process, we explore different noise scheduling strategies from three aspects: noise intensity, noise-adding scheme, and denoising scheme. The noise scheduling of the forward noise-adding process is illustrated in Fig.~\ref{Fig_nas}, while the scheduling of the reverse denoising process is shown in Fig.~\ref{Fig_ds}.

\subsubsection{Noise intensity schedules}
We define two types of noise intensity schedules: Continuous Noise (CN) and Segmented Noise (SN). CN corresponds to the monotonic noise intensity used in canonical diffusion models. The noise strength $\hat\alpha$ varies continuously from 0 to 1 across all timesteps: 
\begin{equation}
\small
    \alpha^{(n)}_\text{CN} = {n}/{N}.
\end{equation}
In DRDM, we need to constrain the intermediate states at different stage nodes of the denoising process to represent mobile traffic with specific spatiotemporal scales. The conventional CN causes all diffusion steps, except for the initial noise-adding point, to contain a certain level of random noise, leading to a deviation between the intermediate states at these nodes and the distributions of traffic at corresponding scales. To address this issue, we propose SN. It linearly schedules $\hat\alpha$ within each stage while maintaining overall continuity in $(0, 1)$:
\begin{equation}
\small
    \alpha_{\text{SN}}^{(n)} = ({n - N_{s_{\delta_k} t_{\tau_k}}})/({\,N_{s_{\delta_{k-1}} t_{\tau_{k-1}}} - N_{s_{\delta_k} t_{\tau_k}}\,}),
\end{equation}
where $n \in \big[N_{s_{\delta_k} t_{\tau_k}},\, N_{s_{\delta_{k-1}} t_{\tau_{k-1}}}\big]$, $k = 1,2,\dots,K$. This strategy ensures that the noise intensity at each intermediate node of the RRDP is zero, allowing noise-free traffic sequences with specific scales to be extracted at the corresponding nodes during the denoising process.

\subsubsection{Noise adding schedules}
We define two types of noise adding schedules: Segmented Noise Adding (SA) and Continuous Noise Adding (CA). SA and CA are independent of the noise intensity scheduling strategy, and their difference lies in whether the starting point of noise addition changes at each stage during the forward process. The CA strategy is consistent with canonical diffusion models, as shown in Eq.~(\ref{eq::forward}). Its noise-adding starting point always remains the data at diffusion step 0, i.e., in Eq.~(\ref{eq::HNaP}), $\pmb{x}^{(0)}_{s_{\delta_k} t_{\tau_k}}=\pmb{x}^{(0)}_{s_{\delta_K} t_{\tau_K}}=\pmb{x}^{(0)}$, 
\begin{equation}
\small
    \alpha^{(n)}_{k,\text{CA}}=\alpha^{(n)},~\alpha^{(n)}\in\{\alpha^{(n)}_\text{CN}, \alpha_{\text{SN}}^{(n)}\}.
\end{equation}
To effectively constrain the intermediate states during the denoising process, we explicitly provide samples at the stage nodes of the noise-adding process. As shown in Fig.~\ref{Fig_nas}, the SA strategy adopts segmented noise addition. Suppose our target is to generate mobile traffic at four spatiotemporal scales $s_1t_1, ~s_2t_2, ~s_3t_3, ~s_4t_4$, ranging from coarse to fine, where $s_4$ represents the smallest spatial scale and $t_4$ denotes the finest temporal resolution. In HNaP, the prior known multi-level mobile traffic $\pmb{x}_{s_4t_4},~\pmb{x}_{s_3t_3},~\pmb{x}_{s_2t_2},~\pmb{x}_{s_1t_1}$ are respectively used as the noise-adding starting points for each stage. That is, in Eq.~(\ref{eq::HNaP}), $\pmb{x}^{(0)}_{s_{\delta_k} t_{\tau_k}} \neq \pmb{x}^{(0)}, ~k \leq K-1$,
\begin{equation}
\small
    \alpha^{(n)}_{k,\text{SA}}=\alpha^{(n-N_{s_{\delta_{k-1}} t_{\tau_{k-1}}})},~\alpha^{(n)}\in\{\alpha^{(n)}_\text{CN}, \alpha_{\text{SN}}^{(n)}\}.
\end{equation}

\subsubsection{Denoising schedules}
Corresponding to the noise adding schedules, we define two types of denoising schedules: Segmented Denoising (SD) and Continuous Denoising (CD). The CD strategy is consistent with canonical diffusion models. For $n \in \{N-1, N-2, ..., 1\}$, the denoising starting point is the Gaussian random noise $\pmb\epsilon$ at $n = N-1$. The SD strategy is a specialized denoising design that works in conjunction with SN. As illustrated in Fig.~(\ref{Fig_ds}), it sets an independent denoising starting point (random noise) for each denoising stage, and each stage individually performs a denoising sub-process from noise to data at the corresponding scale. It is important to note that this is not a cascade of multiple diffusion models, since all stages share the same denoising network $\pmb\epsilon_\theta$.


\subsection{Resolution Greediness Preference in Multi-stage Denoising}
\label{sec::rgp}
\begin{figure*}[tb]
\centering
\includegraphics[width=0.8\linewidth]{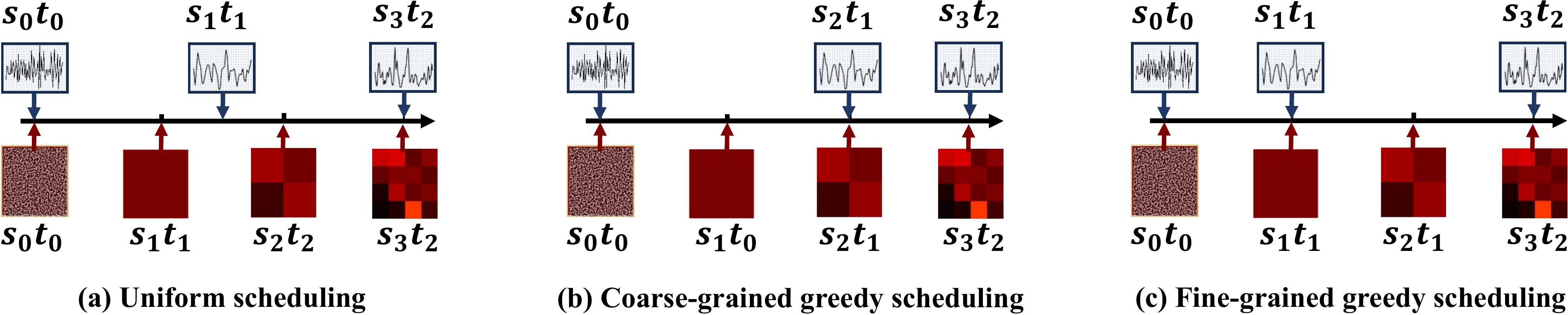}
\caption{Three denoising allocation strategies, including the Uniform Strategy, Coarse-grained Greedy Strategy, and Fine-grained Greedy Strategy.}
\label{Fig_rgp}
\end{figure*}

Resolution Greediness Preference (RGP) refers to the segmentation strategy of the noise-adding and denoising processes. The variation in final generation accuracy under different RGPs reflects the relative importance of precision across resolution stages. We propose three types of RGP as follows: 
\begin{itemize}
    \item \textit{Uniform scheduling}: As shown in Fig.~\ref{Fig_rgp}(a), the diffusion steps are evenly allocated across different spatial scales and temporal resolutions, respectively;
    \item \textit{Coarse-grained greedy scheduling (Coarse Gr.)}: As shown in Fig.~\ref{Fig_rgp}(a), the number of stages $K$ equals the larger of the temporal resolutions and spatial scales. The stages corresponding to the lower spatial or temporal resolution levels are placed toward the later steps of denoising;
    \item \textit{Fine-grained greedy scheduling (Fine Gr.)}: As shown in Fig.~\ref{Fig_rgp}(a), the number of stages $K$ equals the larger of the temporal and spatial resolution levels. The stages corresponding to the higher spatial or temporal resolution levels are placed earlier in the denoising process.
\end{itemize}

\section{Evaluation} 
\label{sec:eval}
To evaluate the performance of ZoomDiff on generating multi-scale mobile traffic with various resolutions, we conduct experiments on a real-world mobile traffic dataset collected from Nanning, China, and compare it against 10 state-of-the-art baselines.

\subsection{Experimental Setup}
\subsubsection{Dataset}
We use a real-world mobile traffic dataset from Nanning, China, which includes mobile traffic records over one week at four layers: BS, Cell, 100m grid, and 50m grid. Due to the limited accessibility of multi-level traffic data, traffic at the BS, Cell, and 100m grid scales is aggregated from the 50m grid traffic according to their respective coverage areas. The finest temporal resolution is 1 hour. Divided by the Yongjiang River, the dataset is split into a South Bank sub-dataset and a North Bank sub-dataset. The South Bank sub-dataset contains 1.4k BSs, 5.7k cells, 23k 100m grids, and 92k 50m grids, while the North Bank subset includes 2.8k BSs, 11k cells, 45k 100m grids, and 182k 50m grids. The dataset is provided by China Mobile.

\subsubsection{Evaluation Metrics}
We use four metrics to evaluate the discrepancy between the generated traffic and the ground-truth traffic. 

\textit{Mean Absolute Error (MAE)} measures the difference between predicted $\hat{\pmb{x}}$ and ground truth $\pmb{x}$: $r_\text{MAE} = \frac{1}{T \cdot H \cdot W} \sum_{t=1}^{T} \sum_{i=1}^{H} \sum_{j=1}^{W} \left| \hat{\pmb{x}}_{t,i,j} - \pmb{x}_{t,i,j} \right|$.

\textit{Root Mean Squared Error (RMSE)} measures the average magnitude of the difference between prediction and ground truth, which penalizes larger deviations more heavily: $r_\text{RMSE} = \sqrt{ \frac{1}{T \cdot H \cdot W} \sum_{t=1}^{T} \sum_{i=1}^{H} \sum_{j=1}^{W} \left( \hat{\pmb{x}}_{t,i,j} - \pmb{x}_{t,i,j} \right)^2 }$.

\textit{Spatial RMSE (spRMSE)} computes RMSE based on time-averaged predictions and ground truths, emphasizing consistency in spatial distribution: $r_\text{spRMSE} = \sqrt{ \frac{1}{H \cdot W} \sum_{i=1}^{H} \sum_{j=1}^{W} \left( \bar{\hat{\pmb{x}}}_{i,j} - \bar{\pmb{x}}_{i,j} \right)^2 }$, where $\bar{\hat{\pmb{x}}}_{i,j} = \frac{1}{T} \sum_{t=1}^{T} \hat{\pmb{x}}_{t,i,j}$, $\bar{\pmb{x}}_{i,j} = \frac{1}{T} \sum_{t=1}^{T} \pmb{x}_{t,i,j}$.

\textit{Peak Signal-to-Noise Ratio (PSNR)} evaluates the reconstruction quality in decibels (dB), using the mean squared error as its basis: $r_\text{PSNR} = 10 \cdot \log_{10} \left( \frac{L^2}{\text{MSE}} \right)$, where $L$ is the maximum possible value of the data.

MAE and RMSE assess the overall accuracy of spatiotemporal sequence generation, while spRMSE and PSNR evaluate the accuracy of spatial distribution.

\subsubsection{Baselines}
We select a total of 10 state-of-the-art baseline models across four categories:  
i) \textbf{Basic forecasting models}: XGBoost~\cite{XGBoost} and Transformer~\cite{Transformer};  
ii) \textbf{Domain-specific models for mobile traffic generation}: KSTDiff~\cite{KSTDiff}, NetDiffus~\cite{NetDiffus}, 5GT-GAN~\cite{5GT-GAN}, and keGAN~\cite{keGAN};
iii) \textbf{Classic diffusion models}: CSDI~\cite{CSDI} and DiT~\cite{DiT}; 
iv) \textbf{Time series generation models}: TimeGAN~\cite{TimeGAN}, and Diffusion-TS~\cite{Diffusion-TS}.

\subsection{Overall Performance}
In this section, we present the overall performance on generating mobile traffic by ZoomDiff and baselines on the South Bank dataset and the North Bank dataset. We conduct experiments on generating traffic at three spatial scales: BS, 100m grid, and 50m grid. The evaluation results on the South Bank dataset are shown in Table~\ref{tab:South-Bank}. The corresponding results on the North Bank dataset are listed in Table~\ref{tab:North-Bank}.
\begin{table*}[htbp]
\footnotesize
\centering
\renewcommand{\arraystretch}{0.95}
\setlength{\tabcolsep}{3.5pt}
\caption{Performance comparison on multi-resolution traffic generation on South Bank, Nanning dataset (\textbf{lower is better} for MAE/RMSE/spRMSE, \textbf{higher is better} for PSNR). Numbers in parentheses indicate variance.  \textbf{Bold} numbers indicate the best-performing model for each metric, and \underline{underlined} numbers indicate the second-best model.}
\begin{tabularx}{\textwidth}{
c|
*{2}{>{\centering\arraybackslash}X}|
*{4}{>{\centering\arraybackslash}X}|
*{4}{>{\centering\arraybackslash}X}
}
\toprule
\multirow{3}{*}{\textbf{Method}} & \multicolumn{2}{c|}{\textbf{BS traffic}} & \multicolumn{4}{c|}{\textbf{100m grid traffic}} & \multicolumn{4}{c}{\textbf{50m grid traffic}} \\
\cmidrule(lr){2-3} \cmidrule(lr){4-7} \cmidrule(lr){8-11}
 & MAE↓ & RMSE↓ & MAE↓ & RMSE↓ & spRMSE↓ & PSNR↑ & MAE↓ & RMSE↓ & spRMSE↓ & PSNR↑ \\
 & e7(e4) & e7(e4) & e7(e4) & e7(e4) & e7(e4) & 1(e-2) & e7(e4) & e7(e4) & e7(e4) & 1(e-2) \\
\midrule
XGBoost       & 1.86\textnormal{\tiny ~(22.7)} & 2.42\textnormal{\tiny ~(36.5)} & 6.31\textnormal{\tiny ~(54.8)} & 7.49\textnormal{\tiny ~(66.2)} & 6.80\textnormal{\tiny ~(61.3)} & 16.1\textnormal{\tiny ~(7.20)} & 8.92\textnormal{\tiny ~(73.5)} & 8.71\textnormal{\tiny ~(84.1)} & 7.97\textnormal{\tiny ~(76.4)} & 13.2\textnormal{\tiny ~(6.70)} \\
Transformer   & 2.02\textnormal{\tiny ~(28.6)} & 2.41\textnormal{\tiny ~(41.0)} & 6.13\textnormal{\tiny ~(50.1)} & 7.13\textnormal{\tiny ~(64.7)} & 6.71\textnormal{\tiny ~(58.9)} & 17.3\textnormal{\tiny ~(7.90)} & 7.41\textnormal{\tiny ~(67.3)} & 7.07\textnormal{\tiny ~(72.8)} & 6.93\textnormal{\tiny ~(69.5)} & 14.8\textnormal{\tiny ~(6.20)} \\

\midrule
KSTDiff       & 1.46\textnormal{\tiny ~(41.4)} & 1.92\textnormal{\tiny ~(47.3)} & 7.77\textnormal{\tiny ~(94.3)} & 12.2\textnormal{\tiny ~(9.61)} & 9.21\textnormal{\tiny ~(11.0)} & 13.0\textnormal{\tiny ~(135)} & 8.76\textnormal{\tiny ~(13.8)} & 16.1\textnormal{\tiny ~(17.2)} & 11.7\textnormal{\tiny ~(16.2)} & 12.9\textnormal{\tiny ~(13.4)} \\
NetDiffus     & 2.36\textnormal{\tiny ~(48.5)} & 2.79\textnormal{\tiny ~(52.2)} & 78.0\textnormal{\tiny ~(98.2)} & 88.4\textnormal{\tiny ~(103)} & 87.5\textnormal{\tiny ~(67.9)} & 9.02\textnormal{\tiny ~(3.63)} & 84.4\textnormal{\tiny ~(76.1)} & 95.6\textnormal{\tiny ~(94.2)} & 101\textnormal{\tiny ~(59.0)} & 8.22\textnormal{\tiny ~(1.29)} \\
5GT-GAN       & 2.63\textnormal{\tiny ~(5.44)} & 3.12\textnormal{\tiny ~(5.48)} & 4.17\textnormal{\tiny ~(7.10)} & 4.97\textnormal{\tiny ~(6.91)} & 5.06\textnormal{\tiny ~(7.09)} & 18.7\textnormal{\tiny ~(2.14)} & 4.50\textnormal{\tiny ~(8.76)} & 6.06\textnormal{\tiny ~(9.84)} & 6.51\textnormal{\tiny ~(9.54)} & 20.8\textnormal{\tiny ~(2.78)} \\
keGAN         & 3.40\textnormal{\tiny ~(99.1)} & 4.58\textnormal{\tiny ~(135)} & 270\textnormal{\tiny ~(341)} & 377\textnormal{\tiny ~(770)} & 317\textnormal{\tiny ~(494)} & 8.29\textnormal{\tiny ~(8.08)} & 377\textnormal{\tiny ~(231)} & 472\textnormal{\tiny ~(631)} & 422\textnormal{\tiny ~(335)} & 6.24\textnormal{\tiny ~(1.27)} \\

\midrule
CSDI          & \underline{1.18}\textnormal{\tiny ~(41.2)} & \underline{1.57}\textnormal{\tiny ~(46.8)} & 2.71\textnormal{\tiny ~(18.2)} & 3.92\textnormal{\tiny ~(22.3)} & 3.69\textnormal{\tiny ~(26.8)} & 26.5\textnormal{\tiny ~(21.9)} & \underline{3.15}\textnormal{\tiny ~(6.76)} & \underline{5.04}\textnormal{\tiny ~(9.00)} & \underline{4.94}\textnormal{\tiny ~(15.8)} & 27.6\textnormal{\tiny ~(22.8)} \\
DIT           & 1.34\textnormal{\tiny ~(28.4)} & 1.76\textnormal{\tiny ~(35.2)} & \underline{2.63}\textnormal{\tiny ~(10.2)} & \underline{3.89}\textnormal{\tiny ~(26.6)} & \underline{3.65}\textnormal{\tiny ~(33.1)} & \underline{26.6}\textnormal{\tiny ~(20.4)} & \underline{3.15}\textnormal{\tiny ~(2.77)} & 5.26\textnormal{\tiny ~(4.13)} & 5.18\textnormal{\tiny ~(6.03)} & \underline{28.1}\textnormal{\tiny ~(4.93)} \\

\midrule
TimeGAN       & 7.00\textnormal{\tiny ~(45.0)} & 6.76\textnormal{\tiny ~(50.3)} & 15.2\textnormal{\tiny ~(66.0)} & 17.0\textnormal{\tiny ~(73.5)} & 16.1\textnormal{\tiny ~(71.2)} & 7.83\textnormal{\tiny ~(3.20)} & 26.4\textnormal{\tiny ~(94.2)} & 28.7\textnormal{\tiny ~(102)} & 27.7\textnormal{\tiny ~(97.1)} & 6.60\textnormal{\tiny ~(2.90)} \\
Diffusion-TS  & 1.35\textnormal{\tiny ~(72.6)} & 1.73\textnormal{\tiny ~(77.8)} & 7.80\textnormal{\tiny ~(0.125)} & 8.82\textnormal{\tiny ~(0.124)} & 9.27\textnormal{\tiny ~(0.223)} & 12.3\textnormal{\tiny ~(0.350)} & 7.95\textnormal{\tiny ~(0.0327)} & 11.1\textnormal{\tiny ~(0.0324)} & 12.1\textnormal{\tiny ~(0.0807)} & 14.4\textnormal{\tiny ~(0.0776)} \\

\midrule
\textbf{ZoomDiff} & \textbf{0.778\textnormal{\tiny ~(7.86)}} & \textbf{1.07\textnormal{\tiny ~(9.23)}} & \textbf{2.17\textnormal{\tiny ~(10.6)}} & \textbf{3.22\textnormal{\tiny ~(10.1)}} & \textbf{2.91\textnormal{\tiny ~(15.5)}} & \textbf{29.1\textnormal{\tiny ~(9.61)}} & \textbf{2.95\textnormal{\tiny ~(9.05)}} & \textbf{4.90\textnormal{\tiny ~(14.4)}} & \textbf{4.72\textnormal{\tiny ~(16.7)}} & \textbf{29.6\textnormal{\tiny ~(21.3)}} \\
\bottomrule
\end{tabularx}
\label{tab:South-Bank}
\end{table*}

\begin{table*}[htbp]
\footnotesize 
\centering
\renewcommand{\arraystretch}{0.95} 
\setlength{\tabcolsep}{3.5pt} 
\caption{Performance comparison on multi-resolution traffic generation on North Bank, Nanning dataset (\textbf{lower is better} for MAE/RMSE/spRMSE, \textbf{higher is better} for PSNR). Numbers in parentheses indicate variance. \textbf{Bold} numbers indicate the best-performing model for each metric, and \underline{underlined} numbers indicate the second-best model.}
\begin{tabularx}{\textwidth}{
c|
*{2}{>{\centering\arraybackslash}X}|
*{4}{>{\centering\arraybackslash}X}|
*{4}{>{\centering\arraybackslash}X}
}
\toprule
\multirow{3}{*}{\textbf{Method}} & \multicolumn{2}{c|}{\textbf{BS traffic}} & \multicolumn{4}{c|}{\textbf{100m grid traffic}} & \multicolumn{4}{c}{\textbf{50m grid traffic}} \\
\cmidrule(lr){2-3} \cmidrule(lr){4-7} \cmidrule(lr){8-11}
 & MAE↓ & RMSE↓ & MAE↓ & RMSE↓ & spRMSE↓ & PSNR↑ & MAE↓ & RMSE↓ & spRMSE↓ & PSNR↑ \\
 & e7(e4) & e7(e4) & e7(e4) & e7(e4) & e7(e4) & 1(e-2) & e7(e4) & e7(e4) & e7(e4) & 1(e-2) \\
\midrule
XGBoost       & 0.960\textnormal{\tiny ~(22.7)} & \underline{1.40}\textnormal{\tiny ~(36.5)} & 4.53\textnormal{\tiny ~(54.8)} & 5.60\textnormal{\tiny ~(66.2)} & 6.27\textnormal{\tiny ~(61.3)} & 20.7\textnormal{\tiny ~(7.2)} & 6.55\textnormal{\tiny ~(73.5)} & 7.45\textnormal{\tiny ~(84.1)} & 7.23\textnormal{\tiny ~(76.4)} & 16.9\textnormal{\tiny ~(6.7)} \\
Transformer   & 2.27\textnormal{\tiny ~(28.6)} & 2.78\textnormal{\tiny ~(41.0)} & 4.57\textnormal{\tiny ~(50.1)} & 5.69\textnormal{\tiny ~(64.7)} & 6.18\textnormal{\tiny ~(58.9)} & 16.1\textnormal{\tiny ~(7.9)} & 6.14\textnormal{\tiny ~(67.3)} & 7.19\textnormal{\tiny ~(72.8)} & 7.64\textnormal{\tiny ~(69.5)} & 17.7\textnormal{\tiny ~(6.2)} \\

\midrule
KSTDiff       & \underline{1.88}\textnormal{\tiny ~(61.8)} & 2.47\textnormal{\tiny ~(78.4)} & 4.55\textnormal{\tiny ~(89.0)} & 6.44\textnormal{\tiny ~(104)} & 6.08\textnormal{\tiny ~(144)} & 23.9\textnormal{\tiny ~(67.2)} & 5.51\textnormal{\tiny ~(34.7)} & 8.69\textnormal{\tiny ~(51.9)} & 8.55\textnormal{\tiny ~(61.2)} & 26.4\textnormal{\tiny ~(34.4)} \\
NetDiffus     & 3.40\textnormal{\tiny ~(32.1)} & 4.91\textnormal{\tiny ~(42.3)} & 21.1\textnormal{\tiny ~(49.8)} & 23.9\textnormal{\tiny ~(63.5)} & 21.5\textnormal{\tiny ~(72.0)} & 10.3\textnormal{\tiny ~(3.62)} & 25.3\textnormal{\tiny ~(43.4)} & 27.9\textnormal{\tiny ~(58.0)} & 32.6\textnormal{\tiny ~(67.4)} & 20.5\textnormal{\tiny ~(2.67)} \\
5GT-GAN       & 2.63\textnormal{\tiny ~(5.44)} & 3.12\textnormal{\tiny ~(5.48)} & 6.53\textnormal{\tiny ~(3.49)} & 7.93\textnormal{\tiny ~(5.00)} & 7.86\textnormal{\tiny ~(4.13)} & 19.1\textnormal{\tiny ~(0.910)} & 10.1\textnormal{\tiny ~(3.23)} & 12.7\textnormal{\tiny ~(4.38)} & 13.0\textnormal{\tiny ~(4.04)} & 21.1\textnormal{\tiny ~(1.19)} \\
keGAN         & 5.29\textnormal{\tiny ~(198)} & 6.76\textnormal{\tiny ~(266)} & 122\textnormal{\tiny ~(128)} & 154\textnormal{\tiny ~(293)} & 134\textnormal{\tiny ~(162)} & 11.1\textnormal{\tiny ~(3.09)} & 172\textnormal{\tiny ~(90.1)} & 233\textnormal{\tiny ~(233)} & 205\textnormal{\tiny ~(152)} & 21.5\textnormal{\tiny ~(4.15)} \\

\midrule
CSDI          & 2.03\textnormal{\tiny ~(43.6)} & 2.68\textnormal{\tiny ~(5.04)} & \underline{3.36}\textnormal{\tiny ~(22.2)} & \underline{4.93}\textnormal{\tiny ~(32.3)} & \underline{4.54}\textnormal{\tiny ~(34.5)} & \underline{27.8}\textnormal{\tiny ~(19.5)} & 5.95\textnormal{\tiny ~(21.5)} & 7.30\textnormal{\tiny ~(13.8)} & \underline{6.50}\textnormal{\tiny ~(12.3)} & 27.4\textnormal{\tiny ~(13.2)} \\
DIT           & 1.97\textnormal{\tiny ~(6.32)} & 2.53\textnormal{\tiny ~(7.43)} & 4.29\textnormal{\tiny ~(25.6)} & 5.73\textnormal{\tiny ~(35.1)} & 5.55\textnormal{\tiny ~(37.2)} & 23.5\textnormal{\tiny ~(13.0)} & \underline{4.13}\textnormal{\tiny ~(5.17)} & \underline{6.83}\textnormal{\tiny ~(8.85)} & 6.75\textnormal{\tiny ~(11.1)} & \underline{28.6}\textnormal{\tiny ~(5.10)} \\

\midrule
TimeGAN & 11.0\textnormal{\tiny ~(45.0)} & 12.4\textnormal{\tiny ~(50.3)} & 24.2\textnormal{\tiny ~(66.0)} & 26.1\textnormal{\tiny ~(73.5)} & 25.2\textnormal{\tiny ~(71.2)} & 6.76\textnormal{\tiny ~(3.20)} & 34.6\textnormal{\tiny ~(94.2)} & 37.4\textnormal{\tiny ~(101)} & 36.2\textnormal{\tiny ~(97.1)} & 7.21\textnormal{\tiny ~(2.90)} \\
Diffusion-TS  & 1.66\textnormal{\tiny ~(38.3)} & 2.31\textnormal{\tiny ~(43.9)} & 6.88\textnormal{\tiny ~(1.71)} & 8.73\textnormal{\tiny ~(2.07)} & 8.17\textnormal{\tiny ~(2.87)} & 16.9\textnormal{\tiny ~(0.550)} & 7.20\textnormal{\tiny ~(2.12)} & 10.9\textnormal{\tiny ~(2.28)} & 9.64\textnormal{\tiny ~(4.90)} & 16.2\textnormal{\tiny ~(0.390)} \\
\midrule
\textbf{ZoomDiff} & \textbf{0.943\textnormal{\tiny ~(7.18)}} & \textbf{1.21\textnormal{\tiny ~(9.25)}} & \textbf{2.23\textnormal{\tiny ~(11.6)}} & \textbf{3.45\textnormal{\tiny ~(13.9)}} & \textbf{3.01\textnormal{\tiny ~(12.0)}} & \textbf{28.9\textnormal{\tiny ~(18.8)}} & \textbf{3.05\textnormal{\tiny ~(9.05)}} & \textbf{4.99\textnormal{\tiny ~(19.1)}} & \textbf{4.87\textnormal{\tiny ~(17.4)}} & \textbf{28.9\textnormal{\tiny ~(31.3)}} \\
\bottomrule
\end{tabularx}
\label{tab:North-Bank}
\end{table*}

According to the results, ZoomDiff performs the best at all resolutions. Compared with the suboptimal model, CSDI, ZoomDiff achieves an average performance improvement of 18.4\%. This demonstrates that ZoomDiff effectively captures the spatiotemporal mapping between the urban environment and network demand. An interesting observation is that, as the spatial resolution increases, diffusion models such as CSDI, DiT and ZoomDiff consistently maintain low generation errors, while GAN-based models like keGAN and TimeGAN suffer from significant performance degradation.  This is because GANs are more prone to mode collapse when faced with increasingly complex spatiotemporal patterns that increase with spatiotemporal resolution. Diffusion models, on the other hand, benefit from the noise-adding and denoising mechanism, leading to strong robustness. Additionally, the progressive denoising process is similar in effect to multi-level mobile traffic with progressive resolution refinement, validating the rationality of employing diffusion models to design resolution-scalable traffic generators.

\subsection{Multi-resolution Generation}
In this section, we compare the denoising intermediate states of CSDI, DiT, and ZoomDiff to validate the effectiveness of our proposed DRDM design. We conduct tests under two resolution configurations:
\begin{itemize}
    \item \textbf{S4T4:} 4 spatial scales (BS, Cell, 100m grid, 50m grid) with 4 temporal resolutions (1 day, 4h, 2h, 1h);
    \item \textbf{S3T4:} 3 spatial scales (Cell, 100m grid, 50m grid) with 4 temporal resolutions (1 day, 4h, 2h, 1h).
\end{itemize}
We adopt a Fine Gr. RGP, where the number of denoising stages equals the greater resolution levels of the temporal dimension and spatial dimension, and the dimension with fewer levels is aligned to earlier denoising steps. For example, under the S3T4 configuration with a total denoising step count of $N = 600$, the stage egment boundarie $\{N_{s_{\delta_k} t_{\tau_k}}\}$ are set at $ \{450, 300, 150\}$. It divides the denoising process into four stages with the following spatiotemporal resolutions: \{(1 day, Cell), (4h, 100m grid), (2h, 50m grid), (1h, 50m grid)\}. For ZoomDiff, we can directly extract the corresponding fine-grained data at $n = 450$, $300$, $150$, and $0$.
\begin{figure}[tb]
\centering
\includegraphics[width=0.95\linewidth]{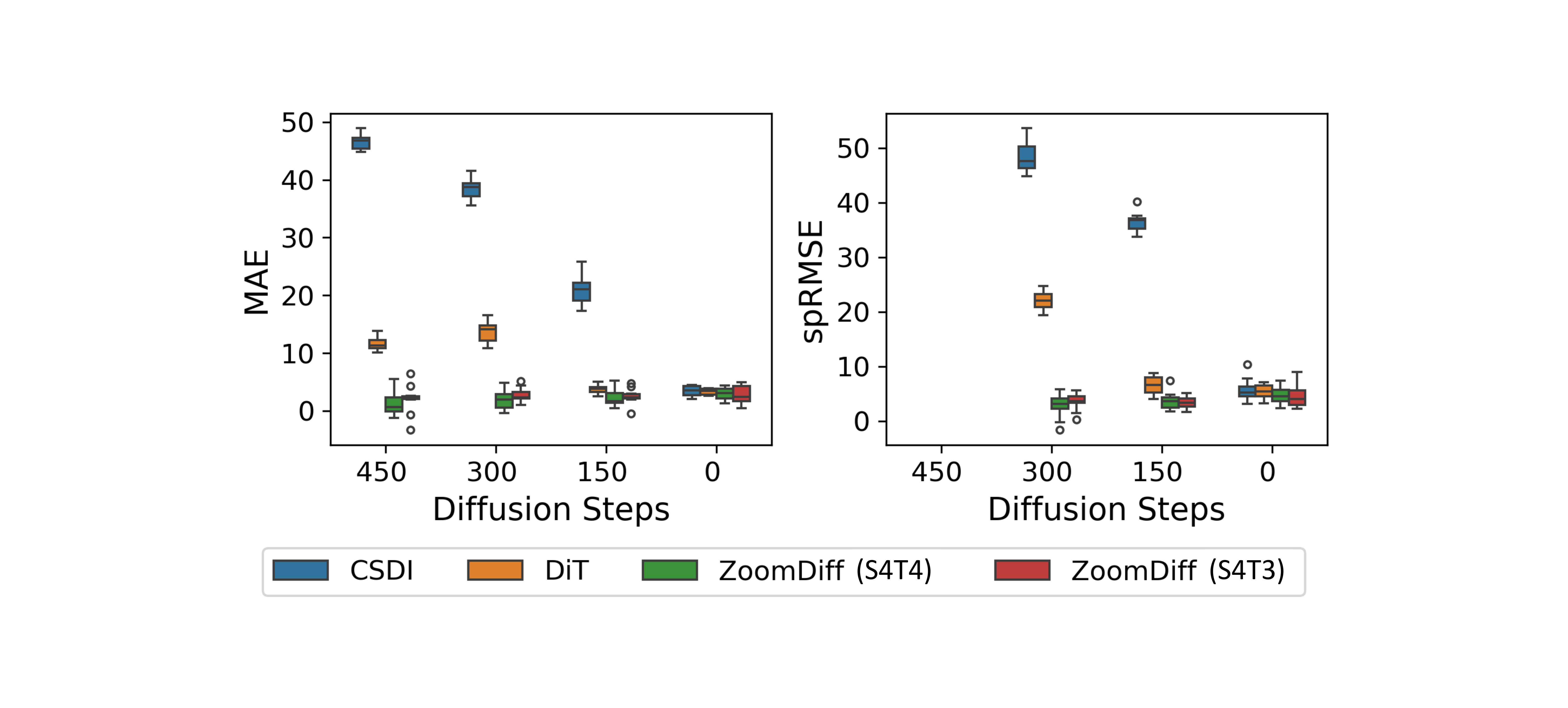}
\caption{Performance comparison of intermediate denoising state across different resolution hierarchy settings for CSDI, DiT, and ZoomDiff on the South Bank, Nanning dataset.}
\label{Fig_box}
\end{figure}

\begin{figure}[tb]
\centering
\includegraphics[width=0.9\linewidth]{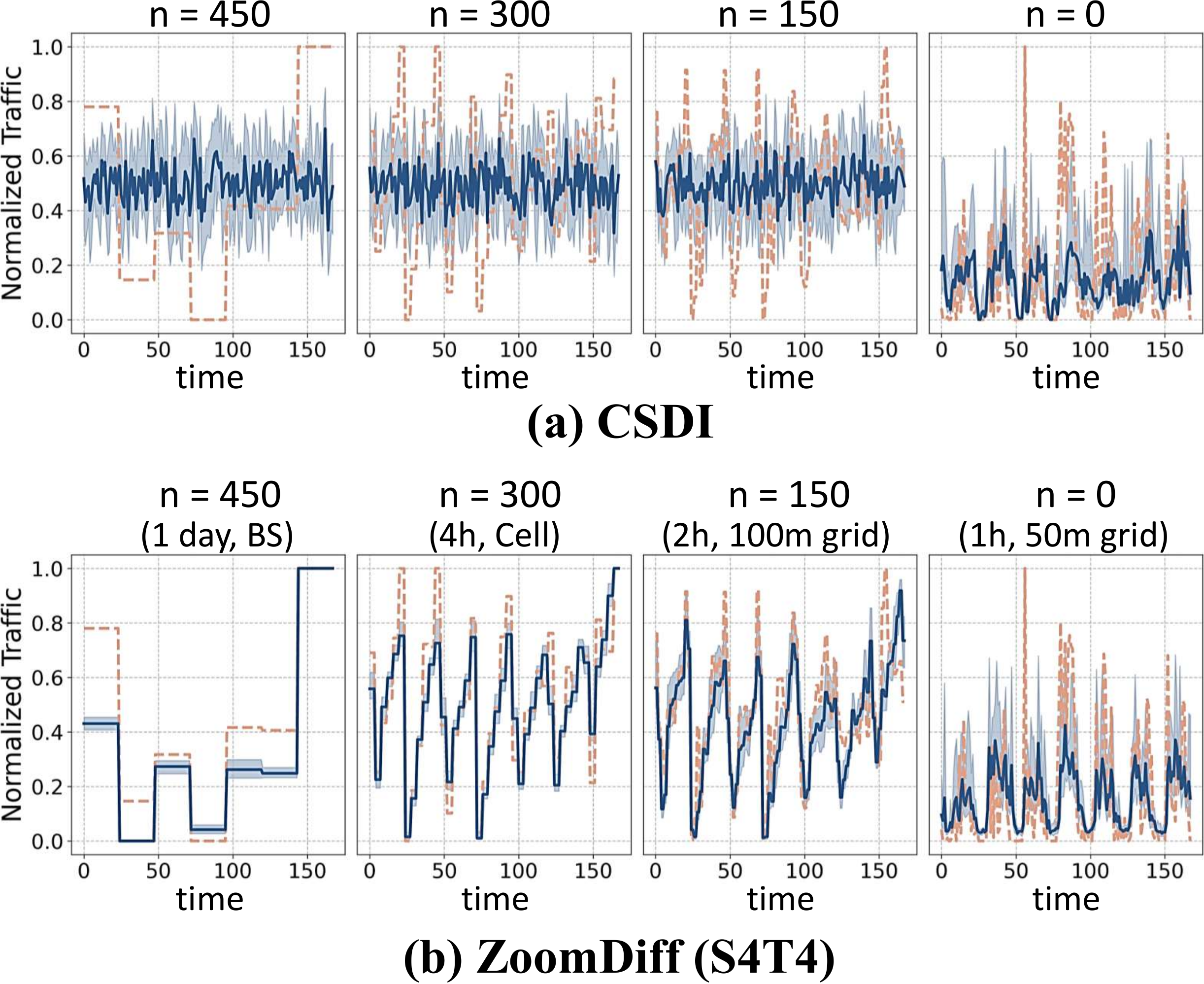}
\caption{Traffic series of intermediate denoising states for CSDI and ZoomDiff (S4T4) on the South Bank, Nanning dataset.}
\label{Fig_time_plot}
\end{figure}

Fig.~\ref{Fig_box} provides a performance comparison of intermediate denoising states across different resolution hierarchy settings for CSDI, DiT, and ZoomDiff on the South Bank, Nanning dataset. To intuitively illustrate the designed RRDP for DRDM, Fig.~\ref{Fig_time_plot} presents the temporal waveforms of intermediate denoising states of CSDI and ZoomDiff (S4T4) extracted at $n=450$, $300$, $150,$ and $0$. 
It can be observed that at $n=0$ (completely denoised), the performances of CSDI, DiT, and ZoomDiff are comparable. However, the intermediate denoising states of CSDI and DiT appear relatively chaotic. In contrast, ZoomDiff is able to produce accurate generation results at different stages of the same denoising process. These intermediate states correspond to mobile traffic with varying spatio-temporal resolutions, rather than noisy and disordered data.

These findings demonstrate that ZoomDiff can effectively regularize the intermediate states of the denoising process without compromising the final high-resolution generation accuracy, which enables the direct extraction of coarser-resolution mobile traffic from intermediate states. This design innovatively overcomes the limitation of canonical diffusion models, which can only generate data at a single resolution, and provides a unified paradigm for multi-resolution mobile traffic generation.

\subsection{Ablation Studies}
\subsubsection{Diffusion Noise Prior Guidance}
\begin{figure}[tb]
\centering
\includegraphics[width=0.85\linewidth]{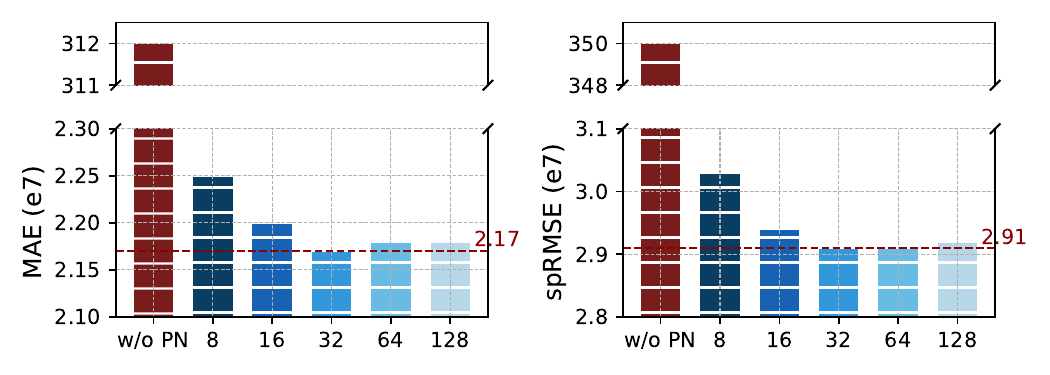}
\caption{Ablation Study: Diffusion Prior Noise. \textit{w/o} PN denotes the removal of prior noise connections between adjacent stages in both the forward and reverse processes. The x-axis values represent the noise dimension $D$.}
\label{Fig_pn}
\end{figure}

As illustrated in Section~\ref{sec::zoomdiff_forward} and Section~\ref{sec::zoomdiff_reverse}, we employ a prior noise guidance to allow low-resolution mobile traffic to guide the generation of high-resolution traffic. To achieve Eq.~\ref{eq::HNaP}, we use a Multi-Layer Perceptron (MLP) to map the predicted residual noise $\pmb\epsilon$ and prior noise $\mathcal{P}_{s_{\delta_{k-1}} t_{\tau_{k-1}}}$ in the $D-$dimensional latent space for concatenation. Fig.~\ref{Fig_pn} shows the impact on high-resolution traffic generation accuracy when removing the prior noise guidance or varying the latent noise dimension $D$. It can be observed that the prior noise guidance is essential for DRDM, as the limited number of diffusion steps in one stage is insufficient to fully learn the complex spatio-temporal patterns of high-resolution traffic. Increasing the latent dimension of the noise helps the model better disentangle and fuse the spatio-temporal patterns embedded in the random and prior noise. Experiments show that the performance converges when $D$ reaches 32, indicating that further increasing the dimension provides no additional information gain in the noise fusion process.

\subsubsection{Spatiotemporal Positional Encoding}
\begin{figure}[tb]
\centering
\includegraphics[width=0.85\linewidth]{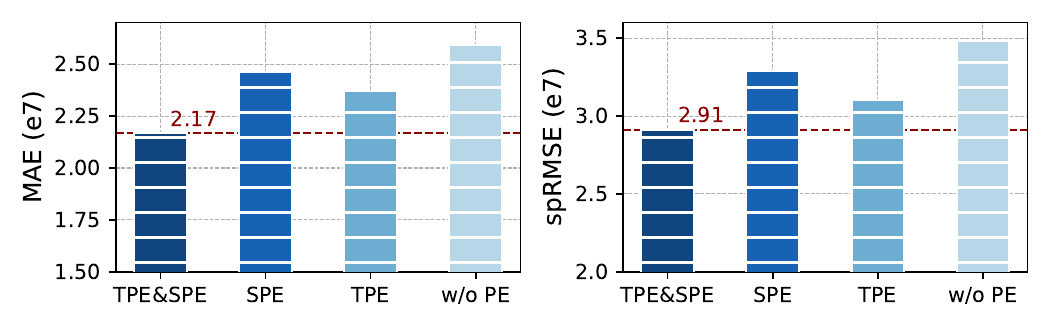}
\caption{Ablation Study: Spatiotemporal Positional Encoding. \textit{TPE\&SPE} indicates the use of both TPE and SPE, \textit{SPE} indicates the use of SPE only, \textit{TPE} indicates the use of TPE only, and \textit{w/o PE} denotes that neither TPE nor SPE is applied.}
\label{Fig_pe}
\end{figure}

In this section, we investigate the role of TPE and SPE in helping the model understand the temporal and spatial characteristics of mobile traffic. We conduct an ablation study targeting this encoding, with the results shown in Fig.~\ref{Fig_pe}. The outcome aligns with our expectations: the introduction of both TPE and SPE provides positive gains to the model.

\subsection{Discussion on Schedualing}
\label{sec::DoS}
\subsubsection{Noise Scheduling}
The performance comparison under different noise scheduling is shown in Table~\ref{tab:noise-scheduling}. There are eight possible combinations of noise intensity scheduling, noise adding scheduling, and denoising scheduling. Among them, [CN, CA, CD] corresponds to the scheduling strategy of canonical diffusion models. We tested five strategy combinations listed in Table~\ref{tab:noise-scheduling}. The combination [CN, *, SD] produced good performance for the final output but resulted in collapse when generating intermediate-scale data. This phenomenon also appeared in the traditional strategy [CN, CA, CD], indicating that CN is not suitable for DRDM’s multi-stage data generation requirements. The [SN, SA, SD] strategy combination achieved the best performance across all resolutions and serves as the optimal noise scheduling strategy adapted for DRDM. Additionally, schedules with SN enable more accurate low-resolution outputs at intermediate denoising stages. These observations indicate that symmetry between the noise-adding and denoising processes is essential for data recovery, and that segmented noise strength scheduling is needed for DRDM. This is because the reduction of intermediate denoising states cannot be achieved simply by adding or subtracting i.i.d. Gaussian noise.
\begin{table*}[htbp]
\footnotesize 
\centering
\renewcommand{\arraystretch}{0.95} 
\setlength{\tabcolsep}{3pt} 
\caption{Comparison of different noise scheduling strategies on South Bank, Nanning dataset (\textbf{lower is better} for MAE/RMSE/spRMSE, \textbf{higher is better} for PSNR). Numbers in parentheses indicate variance.}
\begin{tabularx}{\textwidth}{
c|
*{2}{>{\centering\arraybackslash}X}|
*{4}{>{\centering\arraybackslash}X}|
*{4}{>{\centering\arraybackslash}X}
}
\toprule
\multirow{3}{*}{\textbf{~Noise Sche.~}} & \multicolumn{2}{c|}{\textbf{N=400 (2h, BS)}} & \multicolumn{4}{c|}{\textbf{N=200 (1h, Cell)}} & \multicolumn{4}{c}{\textbf{N=0 (1h, 100m grid)}} \\
\cmidrule(lr){2-3} \cmidrule(lr){4-7} \cmidrule(lr){8-11}
 & MAE↓ & RMSE↓ & MAE↓ & RMSE↓ & spRMSE↓ & PSNR↑ & MAE↓ & RMSE↓ & spRMSE↓ & PSNR↑ \\
 & e7(e4) & e7(e4) & e7(e4) & e7(e4) & e7(e4) & 1(e-2) & e7(e4) & e7(e4) & e7(e4) & 1(e-2) \\
\midrule
CN, CA, CD               & 30.0\textnormal{\tiny ~(38.2)} & 37.4\textnormal{\tiny ~(42.2)} & 17.1\textnormal{\tiny ~(17.9)} & 21.3\textnormal{\tiny ~(16.9)} & 35.3\textnormal{\tiny ~(38.4)} & 9.67\textnormal{\tiny ~(13.2)} & 2.25\textnormal{\tiny ~(12.9)} & 3.32\textnormal{\tiny ~(16.5)} & 2.99\textnormal{\tiny ~(16.6)} & 29.4\textnormal{\tiny ~(24.0)} \\
CN, CA, SD     & 37.1\textnormal{\tiny ~(34.1)} & 4.63\textnormal{\tiny ~(36.9)} & 21.6\textnormal{\tiny ~(24.6)} & 26.9\textnormal{\tiny ~(22.2)} & 25.4\textnormal{\tiny ~(22.1)} & 12.9\textnormal{\tiny ~(10.2)} & 2.95\textnormal{\tiny ~(49.1)} & 4.24\textnormal{\tiny ~(57.6)} & 3.86\textnormal{\tiny ~(70.0)} & 24.5\textnormal{\tiny ~(49.2)} \\
CN, SA, SD     & 40.1\textnormal{\tiny ~(51.9)} & 50.0\textnormal{\tiny ~(73.7)} & 25.1\textnormal{\tiny ~(47.1)} & 31.3\textnormal{\tiny ~(55.3)} & 29.6\textnormal{\tiny ~(54.5)} & 14.1\textnormal{\tiny ~(10.9)} & 2.91\textnormal{\tiny ~(17.0)} & 4.20\textnormal{\tiny ~(23.5)} & 3.79\textnormal{\tiny ~(31.6)} & 26.0\textnormal{\tiny ~(33.5)} \\
SN, SA, CD     & 1.80\textnormal{\tiny ~(24.1)} & 2.13\textnormal{\tiny ~(25.6)} & 4.78\textnormal{\tiny ~(90.0)} & 5.54\textnormal{\tiny ~(98.7)} & 6.04\textnormal{\tiny ~(97.6)} & 10.1\textnormal{\tiny ~(18.8)} & 5.75\textnormal{\tiny ~(126)}  & 7.07\textnormal{\tiny ~(127)}  & 7.73\textnormal{\tiny ~(156)}  & 15.8\textnormal{\tiny ~(65.9)} \\
SN, SA, SD                 & 1.80\textnormal{\tiny ~(24.4)} & 2.12\textnormal{\tiny ~(27.5)} & 2.14\textnormal{\tiny ~(17.4)} & 2.70\textnormal{\tiny ~(18.3)} & 2.60\textnormal{\tiny ~(18.3)} & 19.8\textnormal{\tiny ~(9.36)} & 2.17\textnormal{\tiny ~(10.6)} & 3.22\textnormal{\tiny ~(10.1)} & 2.91\textnormal{\tiny ~(15.5)} & 29.1\textnormal{\tiny ~(9.61)} \\
\bottomrule
\end{tabularx}
\label{tab:noise-scheduling}
\end{table*}

\subsubsection{RGP in Multi-stage Denoising}
We adopt the S3T2 resolution configuration, and the results are shown in Table~\ref{tab:diffusion-scheduling}. The Uniform scheduling divides the denoising process into four stages, while both greedy strategies divide it into three. In terms of the highest-resolution accuracy, the Fine Gr. performs the best, whereas Uniform is relatively less advantageous. This result reflects that, given the same resolution configuration, a smaller number of stages $K$ can lead to more accurate high-resolution generation (with the cost of one fewer observable intermediate denoising state). What's more, sufficient learning of high-resolution spatio-temporal patterns is more critical to achieving high-resolution accuracy than over-emphasizing learning at lower resolutions.
\begin{table*}[htbp]
\footnotesize 
\centering
\renewcommand{\arraystretch}{0.95} 
\setlength{\tabcolsep}{3.5pt} 
\caption{Comparison of different RGP strategies across resolution levels on the South Bank, Nanning dataset.}
\begin{tabularx}{\textwidth}{
c|
*{2}{>{\centering\arraybackslash}X}|
*{4}{>{\centering\arraybackslash}X}|
*{4}{>{\centering\arraybackslash}X}|
*{4}{>{\centering\arraybackslash}X}
}
\toprule
\multirow{2}{*}{\textbf{RGP}} 
& MAE↓ & RMSE↓ & MAE↓ & RMSE↓ & spRMSE↓ & PSNR↑ & MAE↓ & RMSE↓ & spRMSE↓ & PSNR↑ & MAE↓ & RMSE↓ & spRMSE↓ & PSNR↑ \\
& e7(e4) & e7(e4) & e7(e4) & e7(e4) & e7(e4) & 1(e-2) & e7(e4) & e7(e4) & e7(e4) & 1(e-2) & e7(e4) & e7(e4) & e7(e4) & 1(e-2) \\
\midrule
\multirow{2}{*}{Uniform} 
& \multicolumn{2}{c|}{\textbf{n=267 (2h, BS)}}
& \multicolumn{4}{c|}{\textbf{n=200 (1h, Cell)}}
& \multicolumn{4}{c|}{\textbf{n=133 (1h, Cell)}}
& \multicolumn{4}{c}{\textbf{n=0 (1h, 100m grid)}} \\
& 1.84\textnormal{\tiny ~(36.2)} & 2.16\textnormal{\tiny ~(38.6)} 
& 1.82\textnormal{\tiny ~(17.3)} & 2.18\textnormal{\tiny ~(20.0)} & 2.09\textnormal{\tiny ~(2.30)} & 17.2\textnormal{\tiny ~(26.6)} 
& 2.01\textnormal{\tiny ~(17.9)} & 2.50\textnormal{\tiny ~(19.8)} & 2.41\textnormal{\tiny ~(21.9)} & 20.9\textnormal{\tiny ~(29.8)} 
& 2.21\textnormal{\tiny ~(12.7)} & 3.24\textnormal{\tiny ~(16.5)} & 2.95\textnormal{\tiny ~(20.1)} & 28.5\textnormal{\tiny ~(20.4)} \\
\multirow{2}{*}{Coarse Gr.} 
& \multicolumn{2}{c|}{\textbf{n=267 (2h, BS)}}
& \multicolumn{4}{c|}{--} 
& \multicolumn{4}{c|}{\textbf{n=133 (2h, Cell)}}
& \multicolumn{4}{c}{\textbf{n=0 (1h, 100m grid)}} \\
& 1.80\textnormal{\tiny ~(35.9)} & 2.12\textnormal{\tiny ~(37.6)} 
& -- & -- & -- & -- 
& 2.15\textnormal{\tiny ~(16.5)} & 2.69\textnormal{\tiny ~(19.3)} & 2.60\textnormal{\tiny ~(12.4)} & 19.7\textnormal{\tiny ~(32.4)} 
& 2.19\textnormal{\tiny ~(18.8)} & 3.24\textnormal{\tiny ~(18.7)} & 2.92\textnormal{\tiny ~(25.3)} & 28.9\textnormal{\tiny ~(28.1)} \\
\multirow{2}{*}{Fine Gr.} 
& \multicolumn{2}{c|}{\textbf{n=267 (2h, BS)}}
& \multicolumn{4}{c|}{--} 
& \multicolumn{4}{c|}{\textbf{n=133 (1h, Cell)}}
& \multicolumn{4}{c}{\textbf{n=0 (1h, 100m grid)}} \\
& 1.80\textnormal{\tiny ~(24.4)} & 2.12\textnormal{\tiny ~(27.5)} 
& -- & -- & -- & -- 
& 2.14\textnormal{\tiny ~(17.4)} & 2.70\textnormal{\tiny ~(18.3)} & 2.60\textnormal{\tiny ~(18.3)} & 19.8\textnormal{\tiny ~(9.36)} 
& 2.17\textnormal{\tiny ~(10.6)} & 3.22\textnormal{\tiny ~(10.1)} & 2.91\textnormal{\tiny ~(15.5)} & 29.1\textnormal{\tiny ~(9.61)} \\
\bottomrule
\end{tabularx}
\label{tab:diffusion-scheduling}
\end{table*}

\subsection{Efficiency Analysis}
Efficient modeling and generation of spatiotemporal data are crucial for large-scale mobile network applications. ZoomDiff achieves joint generation of multi-scale mobile traffic by producing mobile traffic at multiple layers with different spatial scales and temporal resolutions through a single denoising process of one diffusion model. The time complexity of diffusion models mainly arises from their iterative denoising steps. Traditional diffusion models can only generate data at a single spatiotemporal resolution, leading to redundant computations when multi-scale data are required. In contrast, ZoomDiff enhances the utilization efficiency of the denoising process by enabling multiple resolutions to be generated simultaneously within a single inference cycle, thereby significantly improving computational efficiency.

To verify the efficiency advantage of ZoomDiff, we conduct a four-level mobile traffic generation task (BS, cell, 100m grid, 50m grid) for an area containing 500 BSs, including both pre-training and inference. For simplicity, the temporal resolution of traffic at each level is fixed at 1 hour. The batch size is set to 4, and a single NVIDIA A100 GPU is used. Table~\ref{tab:efficiency} presents the comparison results between ZoomDiff and representative diffusion models (CSDI, DIT) in terms of parameter scale, training speed, and inference speed. It can be observed that ZoomDiff, while maintaining a small number of parameters (0.504M), achieves remarkable acceleration performance. Its training time is reduced from 22.9 s/epoch in CSDI to 12.25 s/epoch, representing a 46.5\% improvement in training efficiency; and its inference time is reduced from 896.6 s/epoch in CSDI to 446.2 s/epoch, representing a 50.2\% improvement in inference efficiency. These results demonstrate the superior efficiency of ZoomDiff when generating multi-scale mobile traffic.
\begin{table}[htbp]
\footnotesize
\centering
\renewcommand{\arraystretch}{0.95}
\setlength{\tabcolsep}{3.5pt}
\caption{Comparison of model efficiency.}
\begin{tabularx}{0.46\textwidth}{
c|
>{\centering\arraybackslash}X|
>{\centering\arraybackslash}X|
>{\centering\arraybackslash}X
}
\toprule
\multirow{2}{*}{\textbf{Method}} & \multirow{2}{*}{\textbf{Para. Size (M)}}
& \textbf{Training Speed} & \textbf{Inference Speed} \\
&                & (s/epoch)      & (s/epoch) \\
\midrule
CSDI     & 0.748 & 22.9  & 896.6 \\
DIT      & 0.99  & 146   & 1001  \\
ZoomDiff     & 0.504 & 12.25 & 446.2 \\
\bottomrule
\end{tabularx}
\label{tab:efficiency}
\end{table}

\section{Case Study}
\label{sec:dis}
In this section, we discuss two key aspects of ZoomDiff: its capability for traffic refinement from coarse (BS level) to fine granularity, and its generalization ability across different cities and network environments.

\subsection{Refinement of BS-level Traffic Data} 
\label{sec:apply}
The refinement of coarse-grained traffic into finer resolutions is a crucial task for enhancing the interpretability and usability of mobile network data. In real-world scenarios, operators typically obtain only large-scale traffic data such as BS-level or cell-level measurements, while finer-grained traffic at the grid level often depends on point sampling or drive testing, which are costly and prone to errors. 
Leveraging its understanding of multi-scale correlations, the pre-trained ZoomDiff model can refine BS or cell-level traffic into grid-level traffic through a zero-shot inference process. Specifically, by initializing the reverse diffusion process with coarse-scale traffic as the denoising starting point, ZoomDiff can progressively generate finer-grained spatial distributions without additional retraining. Table~\ref{tab:apply} presents the comparison results for refining BS traffic into 50m-grid traffic using different approaches. The results show that \textit{ZoomDiff (zero-shot)} achieves accuracy close to that of retrained models, significantly outperforming traditional baselines such as CSDI. This demonstrates that ZoomDiff effectively captures the hierarchical correlations in mobile traffic and supports efficient fine-grained refinement.
\begin{table}[htbp]
\footnotesize
\centering
\renewcommand{\arraystretch}{0.95}
\setlength{\tabcolsep}{3.5pt}
\caption{Comparison of BS traffic refinement.}
\begin{tabularx}{0.46\textwidth}{
c|
*{4}{>{\centering\arraybackslash}X}
}
\toprule
\multirow{2}{*}{\textbf{Method}} 
& MAE↓ & RMSE↓ & spRMSE↓ & PSNR↑ \\
& e7(e4) & e7(e4) & e7(e4) & 1(e-2) \\
\midrule
CSDI & 3.47\textnormal{\tiny ~(12.7)} & 5.65\textnormal{\tiny ~(16.5)} & 5.57\textnormal{\tiny ~(20.1)} & 23.9\textnormal{\tiny ~(20.4)} \\
ZoomDiff (zero-shot) & 3.54\textnormal{\tiny ~(7.42)} & 5.44\textnormal{\tiny ~(9.36)} & 5.67\textnormal{\tiny ~(15.5)} & 23.3\textnormal{\tiny ~(6.59)} \\
ZoomDiff & 3.24\textnormal{\tiny ~(5.02)} & 5.21\textnormal{\tiny ~(5.11)} & 5.17\textnormal{\tiny ~(7.99)} & 24.6\textnormal{\tiny ~(6.44)} \\
\bottomrule
\end{tabularx}
\label{tab:apply}
\vspace{-4mm}
\end{table}

\subsection{Generalization Validation}
Mobile traffic patterns often differ significantly across cities due to variations in population density, infrastructure deployment, and user behavior. However, it is impractical to collect sufficient training data for every city due to privacy and deployment constraints. Therefore, the ability to generalize learned traffic generation models across cities is essential. To evaluate the cross-city generalization of ZoomDiff, we conducted transfer experiments between Nanning and Shanghai using Macro-BS (MBS) and Pico-BS (PBS) datasets. In the transfer setting, models trained on Nanning data were tested on Shanghai data without fine-tuning. The results, shown in Table~\ref{tab:trans}, indicate that ZoomDiff achieves consistently better performance than CSDI in both single-city training and cross-city transfer. Notably, the degradation in accuracy during transfer is minor, demonstrating that ZoomDiff effectively captures universal spatiotemporal dependencies that are transferable across diverse urban environments.
\begin{table}[htbp]
\footnotesize
\centering
\renewcommand{\arraystretch}{0.95}
\setlength{\tabcolsep}{3.5pt}
\caption{Comparison of model transfer performance for PBS traffic generation. Nanning$\rightarrow$Shanghai means training models on the Nanning dataset and testing models on the Shanghai dataset.}
\begin{tabularx}{0.46\textwidth}{
c|
c|
>{\centering\arraybackslash}X|
>{\centering\arraybackslash}X|
>{\centering\arraybackslash}X
}
\toprule
\multirow{2}{*}{\textbf{Method}} & \multirow{2}{*}{\textbf{Dataset}} 
& \textbf{MAE↓} & \textbf{RMSE↓} & \textbf{PSNR↑} \\
& & \small e3(e1) & \small e3(e1) & \small 1(e-1) \\
\midrule
\multirow{3}{*}{CSDI} 
& Nanning & 1.01\tiny{~(1.76)} & 1.28\tiny{~(1.93)} & 19.31\tiny{~(2.10)} \\
& Shanghai & 1.88\tiny{~(1.47)} & 2.05\tiny{~(2.25)} & 29.11\tiny{~(2.46)} \\
& Nanning$\rightarrow$Shanghai & 2.42\tiny{~(2.54)} & 2.87\tiny{~(2.93)} & 26.50\tiny{~(2.91)} \\
\midrule
\multirow{3}{*}{\textbf{ZoomDiff}} 
& Nanning & \textbf{0.610}\tiny{~(0.946)} & \textbf{0.719}\tiny{~(0.942)} & \textbf{26.0}\tiny{~(2.89)} \\
& Shanghai & \textbf{1.63}\tiny{~(2.49)} & \textbf{1.95}\tiny{~(2.61)} & \textbf{37.1}\tiny{~(1.36)} \\
& Nanning$\rightarrow$Shanghai & \textbf{2.26}\tiny{~(1.08)} & \textbf{2.58}\tiny{~(1.04)} & \textbf{33.3}\tiny{~(1.49)} \\
\bottomrule
\end{tabularx}
\label{tab:trans}
\end{table}

\section{Conclusion} 
\label{sec:con}
In this paper, we propose \textbf{ZoomDiff}, a diffusion-based model for multi-scale mobile traffic generation. ZoomDiff employs customized multi-resolution diffusion models, DRDM. It achieves joint modeling of multi-layer mobile traffic with different spatial and temporal resolutions through multi-stage noise-adding and denoising processes. By leveraging a prior noise guidance mechanism and hierarchical denoising stages, ZoomDiff can generate multi-scale mobile traffic within a single denoising process by controlling intermediate states. Furthermore, we investigate effective noise scheduling strategies and analyze the resolution greediness preference in DRDM, which enhances the stability and scalability of the model. Extensive experiments on real-world datasets demonstrate that ZoomDiff achieves at least an 18.4\% improvement over state-of-the-art baselines in generating multi-scale mobile traffic and shows strong cross-city generalization capability. These results highlight ZoomDiff as a promising foundation for future research in generative modeling of large-scale mobile network dynamics.


\bibliographystyle{IEEEtran}
\bibliography{Ref}
\vfill

\end{document}